\documentclass[10pt,aps,prd,floatfix,notitlepage,nofootinbib]{revtex4-1}
\usepackage[utf8]{inputenc}
\usepackage{multirow}
\usepackage{latexsym}
\usepackage{epsfig,graphics,subfig,float}

\usepackage{tabularx,booktabs,caption}% http://ctan.org/pkg/{tabularx,booktabs,caption}

\usepackage{amssymb}
\usepackage{verbatim}
\usepackage{color}
\usepackage{tikz}
\usepackage{amsmath,amssymb,amsfonts,hyperref}

\usepackage{epstopdf}
\usepackage{graphicx}
\usepackage{comment}
\usepackage{soul}

\usepackage{tikz}
\usepackage{bm}
\usetikzlibrary{matrix}

\begin{document}

\title{New Exact Spherically Symmetric Solutions in $f(R,\phi,X)$ gravity by Noether's symmetry approach}

\author{Sebastian Bahamonde$^{1,2,3,4}$, Kazuharu Bamba$^{5}$ and Ugur Camci$^{6}$}
\email{$^{1,2,3}$ sebastian.beltran.14@ucl.ac.uk, sbahamonde@ut.ee \\
$^{4}$ bamba@sss.fukushima-u.ac.jp \\$^{5}$ ugurcamci@gmail.com }

\affiliation{$^{1}$Laboratory of Theoretical Physics, Institute of Physics, University of Tartu, W. Ostwaldi 1, 50411 Tartu, Estonia\\
$^{2}$Department of Mathematics, University College London, Gower Street, London, WC1E 6BT, United Kingdom \\
$^{3}$School of Mathematics and Physics, University of Lincoln.	Brayford Pool, Lincoln, LN6 7TS, United Kingdom\\
$^{4}$University of Cambridge, Cavendish Laboratory, JJ Thomson Avenue, Cambridge CB3 0HE, United Kingdom\\
$^{5}$Division of Human Support System, Faculty of Symbiotic Systems Science, Fukushima University, Fukushima 960-1296, Japan\\
$^{6}$Siteler Mahallesi, 1307 Sokak, Ahmet Kartal Konutlari, A-1
Blok, No:7/2, 07070, Konyaalti, Antalya, Turkey
}

\begin{abstract}
The exact solutions of spherically symmetric space-times are explored by using Noether symmetries in $f(R,\phi,X)$ gravity with $R$ the scalar curvature, $\phi$ a scalar field and $X$ the kinetic term of $\phi$. Some of these solutions can represent new black holes solutions in this extended theory of gravity. The classical Noether approach is particularly applied to acquire the Noether symmetry in $f(R,\phi,X)$ gravity. Under the classical Noether theorem, it is shown that the Noether symmetry in $f(R,\phi,X)$ gravity yields the solvable first integral of motion. With the conservation relation obtained from the Noether symmetry, the exact solutions for the field equations can be found. The most important result in this paper is that, without assuming $R=\textrm{constant}$, we have found new spherically symmetric solutions in different theories such as: power-law $f(R)=f_0 R^n$ gravity, non-minimally coupling models between the scalar field and the Ricci scalar $f(R,\phi,X)=f_0 R^n \phi^m+f_1 X^q-V(\phi)$, non-minimally couplings between the scalar field and a kinetic term $f(R,\phi,X)=f_0 R^n +f_1\phi^mX^q$ , and also in extended Brans-Dicke gravity $f(R,\phi,X)=U(\phi,X)R$. It is also demonstrated that the approach with Noether symmetries can be regarded as a selection rule to determine the potential $V(\phi)$ for $\phi$, included in some class of the theories of $f(R,\phi,X)$ gravity.
\end{abstract}

\pacs{04.30, 04.30.Nk, 04.50.+h, 98.70.Vc}

%\date{\today}

\maketitle
%\tableofcontents

\section{Introduction}\label{sec:0}

Through the current observations such as Type Ia Supernovae~\cite{SN},
cosmic microwave background (CMB) radiation~\cite{CMB},
large scale structure~\cite{LSS},
baryon acoustic oscillations (BAO)~\cite{Eisenstein:2005su} and weak lensing~\cite{Jain:2003tba},
it has been revealed that the cosmic expansion is accelerating
at the present time as well as in the early universe at the inflationary
stage~\cite{Inflation}.
Two main approaches have been proposed in order to account for
the late-time accelerated expansion of the universe.
The first is the way of assuming the existence of dark energy
in the framework of general relativity.
The second is that of modifying the theories of gravitation
at the large-scales (for recent reviews on the theories of modified gravity and the issue of dark energy, see, for instance,~\cite{M-G-Reviews}).

In addition, more recently, LIGO has detected that in the coalescence phase,
strong gravitational waves are emitted by the system of two black holes~\cite{Abbott:2016blz}. The first event was the emission from black holes whose masses
are about 30 solar ones and the following ones are those from some-black-holes mergers~\cite{Abbott:2016nmj, Abbott:2017vtc, Abbott:2017oio, Abbott:2017gyy}. Furthermore, strong gravitational waves have been discovered from the two-neutron-stars collision~\cite{TheLIGOScientific:2017qsa},
and this fact led to the multi messenger astronomy.

For $f(R)$ gravity and the scalar-tensor theories~\cite{Nojiri:2017hai},
with the Neutron Star Merger GW170817~\cite{{TheLIGOScientific:2017qsa}},
the cosmological bounds have been studied. Also, by using GW150914 and GW151226~\cite{TheLIGOScientific:2016src, Abbott:2016nmj, TheLIGOScientific:2016pea},
the observational constraints on modified gravity theories have been examined~\cite{DeLaurentis:2016jfs}. Gravitational waves in the context of modified gravity theories have been analyzed~\cite{Capozziello:2007zza, Capozziello:2008fn, Capozziello:2008rq, Bellucci:2008jt, Bogdanos:2009tn, Capozziello:2017vdi,Bamba:2018cup}.
Under such current situations, it is very significant to investigate the solutions of spherically symmetric solutions that could describe black holes, which are the sources emitting gravitational waves, in modified gravity theories in detail so that we can find some clues to deduce information of fundamental physics in strong gravity regions.

The various laws of conservation such that energy conservation, momentum conservation, angular momentum conservation, etc., are directly related with symmetries of a given dynamical system and provide the integrals of motion which are indeed the result of existence of some kinds of symmetry in that system. Using the Noether Symmetry Approach, it is possible to obtain conserved quantities asking for the symmetries of the Lagrangian. The existence of some kinds of symmetry for the Euler-Lagrange equations of motion possessing a Lagrangian would immediately be connected with the Noether symmetry. Even if there is no any specific theory favored by the Noether symmetry approach, the discussions from the literature \cite{noether} point out that the existence of Noether symmetries is capable of selecting suitable gravity theory and then to integrate dynamics by using the first integrals corresponding to the Noether symmetries. A consequent process with regard to the first integrals due to the Noether symmetries allows to achieve exact solutions of the dynamical equations for the gravity theory. Furthermore, it should be noted that the Noether symmetries are not only a mathematical tool to solve or reduce dynamics but also their existence allows to select observable universes/black holes/wormholes etc. and to select analytical models related to observations \cite{capo2012}. The existence of a black hole (or any astrophysical object) is due to the solution of field equations for the selected theory of gravity, which provides a non-trivial structured linkage of disappeared points in the space-time. In particular, it is possible to classify singularity behaviors of any gravity theory which may be related to Noether symmetries, and then decide which one is physical or unphysical solution.

In this paper, with Noether symmetries,
we investigate the exact solutions of spherically symmetric spacetimes in $f(R,\phi,X)$ gravity,
where $R$ is the scalar curvature, $\phi$ is a scalar field, $X$ is
the kinetic term of $\phi$, and $f(R,\phi,X)$ is a function of $R$, $\phi$ and $X$ \cite{Beltran:2015hja,Zubair:2017oir}. This theory can describe various modified gravity theories including the scalar-tensor gravity and $f(R)$ gravity.
In particular, we adopt the classical Noether approach in order to find the Noether symmetry in $f(R,\phi,X)$ gravity. See \cite{noether} for some studies related to Noether symmetry approach in modified gravity. As a result, from the classical Noether theorem, it is shown that the Noether symmetry in $f(R,\phi,X)$ gravity leads to a kind of the first integral of motion, which are able to be solved, so that we need not to analyze the cyclic coordinate, as explored in detail in Ref.~\cite{capo2007}. Thus, we derive exact solutions for the field equations by using
the conservation relation coming from the Noether symmetry acquired.
Moreover, it is demonstrated that the approach with Noether symmetries can be regarded as a selection rule to determine the potential form $V(\phi)$ of $\phi$, which exists in some class of theories described as $f(R,\phi,X)$ gravity. One important approach in this paper will be to obtain new spherically symmetric solutions in this extended theory of gravity without assuming $R=\textrm{constant}$, as in other papers. If one assumes $R=\textrm{constant}$, one looses the higher order terms coming from $f(R)$ gravity, making the theory not so interesting.
%%% Organization %%%
The organization of the present paper is as follows.
In Sec.~II, we explain $f(R,\phi,X)$ gravity in spherically symmetric space-time.
In Sec.~III, we explore the symmetry reduced Lagrangian in $f(R,\phi,X)$ gravity.
In Sec.~IV, we investigate the approach with the Noether symmetry.
In Sec.~V, conclusions are finally presented.
%%%%%%%%%%%%%%%%%%%%

%%%%%%%%%%%%%%%%%%%%%%%%%%%
%%%  Sec. II
%%%%%%%%%%%%%%%%%%%%%%%%%%%
\section{$f(R,\phi,X)$ gravity in spherically symmetric space-time}
The action that we will consider reads~\cite{Beltran:2015hja}
\begin{equation}
S=\int d^4x \sqrt{-g}\left[ \frac{1}{2\kappa^2} f(R,\phi,X) + L_{\rm m} \right]\,,
\label{action}
\end{equation}
where $\kappa^2=8\pi G$, $L_{\rm m}$ is any matter Lagrangian and $f$ is a function which depends on the scalar curvature $R$, a scalar field $\phi$ and a kinetic term being equal to
\begin{eqnarray}
X=-\frac{\epsilon}{2}\, \partial^{\mu}\phi\partial_{\mu}\phi\,,
\end{eqnarray}
where $\epsilon$ is a parameter that if is equal to 1 represents canonical scalar field and equal to $-1$ represents a phantom scalar field. Clearly, the above action has many different scalar tensor theories such as Brans-Dicke types ($f(R,\phi,X)=\gamma(\phi,X)R$) or minimally coupled scalar tensor theories  ($f(R,\phi,X)=\alpha(R)+\gamma(\phi,X)$). Variations of the action \eqref{action} with respect to the metric yields
\begin{equation}
f_{R} G_{\mu\nu}=\frac{1}{2}\left( f-Rf_{R}  \right) g_{\mu\nu} + \nabla_\nu \nabla_{\mu} f_{R}- g_{\mu \nu}\nabla_\alpha \nabla^\alpha f_{R} + \frac{\epsilon}{2}f_X \, (\nabla_\mu\phi) (\nabla_\nu \phi)\,,
\label{fieldeq1}
\end{equation}
whereas variations with respect to the scalar field $\phi$ gives 
\begin{equation}
\nabla_\mu \left( f_X\,\nabla^\mu\phi  \right) + \epsilon f_\phi =0 \,.
\label{fieldeq2}
\end{equation}
Here, we have assumed the vacuum case where $L_{\rm m}=0$ and $f_R=\partial f/\partial R$, $f_X=\partial f/\partial X$ and $f_\phi=\partial f/\partial \phi$.
It should be noted that the Schwarzschild solution is the unique spherically symmetric vacuum solution in GR, but we will see that this no longer holds in $f(R,\phi,X)$ theory of gravity. We also mention here that the vacuum solutions do not necessarily imply a null curvature $R=0$ or $R=const.$, which lead to maximally symmetric solutions, to the contrary in GR. The trace of Eq.(\ref{fieldeq1}) yields
\begin{equation}
R f_R + X f_X - 2 f(R,\phi,X) + 3 \square f_R = 0,  \label{trace}
\end{equation}
where $\square = \nabla^{\mu} \nabla_{\mu}$. The latter equation will be useful in studying various aspects of $f(R,\phi,X)$ gravity.

Let us now consider that the space-time is spherically symmetric such as the metric is
\begin{eqnarray}
ds^2=-A(r)dt^2+B(r)dr^2+M(r)(d\theta^2 + \sin\theta^2 d\varphi^2), \label{metric}
\end{eqnarray}
where $A(r), B(r)$ and $M(r)$ are functions of the radial coordinate $r$. It is easy to see that in this space-time, the scalar curvature is
\begin{eqnarray}
R = -\frac{1}{B} \left[  \frac{A''}{A} + \frac{ 2M''}{M}  - \frac{A' B'}{2 A B} + \frac{M' A'}{M A} - \frac{M' B'}{M B} - \frac{A'^2}{2 A^2} - \frac{M'^2}{2 M^2} -  \frac{2 B}{M}  \right]. \label{r-scalar}
\end{eqnarray}
For the metric \eqref{metric}, the field equations \eqref{fieldeq1} become
\begin{equation}
f_R \left( 2 \frac{A''}{A} - \frac{A' B'}{A B} - \frac{A'^2}{A^2} + 2 \frac{M' A'}{M A} \right) = - 2 B f + 4 \left[ f''_R + f'_R \left( \frac{M'}{M} - \frac{B'}{2 B} \right)  \right], \label{feq1}
\end{equation}
\begin{equation}
f_R \left( 2 \frac{A''}{A} + 4 \frac{M''}{M} - \frac{A' B'}{A B} - 2 \frac{M' B'}{M B} - \frac{A'^2}{A^2} - 2 \frac{M'^2}{M^2}  \right) = - 2 B f + 2 f'_R \left( \frac{A'}{A} + 2 \frac{M'}{M} \right) - 2 \epsilon f_X \phi'^2, \label{feq2}
\end{equation}
\begin{equation}
f_R \left( 2 \frac{M''}{M} - \frac{B' M'}{B M} +  \frac{M' A'}{M A} - \frac{4 B}{M} \right) = - 2 B f + 4 \left[ f''_R + \frac{1}{2} f'_R \left( \frac{A'}{A} -\frac{B'}{B} + \frac{M'}{M} \right)  \right], \label{feq3}
\end{equation}
whereas the modified Klein-Gordon equation \eqref{fieldeq2} yields
\begin{eqnarray}
f_X \left[ \phi'' + \frac{1}{2} \phi' \left( \frac{A'}{A} - \frac{B'}{B} + 2 \frac{M'}{M} \right)  \right] + f'_X \phi'  + \epsilon B f_{\phi} =0. \label{kgeq}
\end{eqnarray}
In above equations, primes denote differentiation with respect to $r$, therefore, the terms $f'_{R}=f_{RR}R' + f_{R\phi}\phi' + f_{RX} X'$ and $f'_{X}=f_{RX} R' + f_{X\phi}\phi' + f_{XX}X'$.

%%%%%%%%%%%%%%%%%%%%%%%%%%%
%%%  Sec. III
%%%%%%%%%%%%%%%%%%%%%%%%%%%
\section{Symmetry reduce Lagrangian in  $f(R,\phi,X)$ gravity}

For simplicity let us express the scalar curvature as follows
\begin{eqnarray}
\bar{R} = R^{*} - \frac{A''}{AB} - \frac{2M''}{BM}\,,  \label{R}
\end{eqnarray}
where
\begin{equation}
R^{*}=\frac{A' B'}{2 A B^2} -\frac{A' M'}{A B M}+\frac{A'^2}{2 A^2 B}+\frac{B' M'}{B^2 M}+\frac{M'^2}{2 B M^2}+\frac{2}{M}
\end{equation}
contains only first derivatives terms. One can rewrite the action into its canonical form in such a way that we can reduce the number of degrees of freedom. In our case, we have
\begin{equation}
S_{f(R,\phi,X)}=\int dr\mathcal{L}(A,A',B,B',M,M',R,R',\phi,\phi') \,,  \label{actioncan0}
\end{equation}
Recall that $X$ depends on $\phi'$. Then, the action \eqref{action} in spherically symmetric space-time \eqref{metric} becomes
\begin{equation}
S_{f(R,\phi,X)}= \int dr\left\{ f(R,\phi,X)-\lambda_{1} \left( R - \bar{R} \right) - \lambda_{2} \left( X - \bar{X} \right) \right\} M \sqrt{AB} \label{actioncan}
\end{equation}
where $\bar{X} = - \frac{\epsilon}{2B}\, \phi'^2$. Here $\lambda_1$ and $\lambda_2$ are the Lagrangian multipliers that can be directly found by varying with respect to $R$ and $X$, giving $\lambda_1=f_{R}$ and $\lambda_2=f_{X}$ respectively. Then, the above canonical action can be rewritten as
\begin{eqnarray}
S_{f(R,\phi,X)}&=& \int dr\left\{f(R,\phi,X)-f_{R} \left[ R - \Big(R^{*}-\frac{A''}{AB}-\frac{2M''}{BM}\Big)\right] -f_{X} \left( X + \frac{\epsilon}{2B}\, \phi'^2\right)\right\} M \sqrt{AB} \,,  \label{actioncan2}\\
&=& \int dr\Big\{M \sqrt{AB} \Big[f(R,\phi,X)-f_{R}(R - R^{*})\Big] + 2 M' \Big(\sqrt{\frac{A}{B}}f_{R}\Big)' + A'\Big(\frac{Mf_R}{\sqrt{AB}}\Big)' \nonumber\\
&& \qquad \quad  -M \sqrt{AB} f_{X} \left( X + \frac{\epsilon}{2B}\, \phi'^2\right)\Big\} \,,  \label{actioncan3}
\end{eqnarray}
where we have integrated by parts and ignored boundary terms. Then, the symmetry reduced Lagrangian becomes
\begin{eqnarray}
\mathcal{L}_f= \sqrt{A B} \left[ M ( f -X f_X) + (2- M R) f_R \right] + \frac{f_R M' A'}{\sqrt{A B}} + \frac{1}{2} f_R  \sqrt{\frac{A}{B}}\frac{M'^2}{ M} + \frac{M A' f'_R}{\sqrt{AB}}  + 2\sqrt{\frac{A}{B}} f'_R M'  - \frac{\epsilon}{2} f_X  M \sqrt{\frac{A}{B}} \phi'^2\,. \qquad \label{lagr}
\end{eqnarray}
Note again that $f_R'=f_{RR}R'+f_{R\phi}\phi'+f_{RX}X'$. Varying the symmetry reduced Lagrangian (\ref{lagr}) with respect to the metric coefficients $A, B, M$, and the scalar field $\phi$ we obtain, respectively
\begin{equation}
f_R \left[ \frac{2 M''}{M} - \frac{M' B'}{M B} - \frac{M'^2}{2 M^2} - B \left( \frac{2}{M} -R \right) \right] + f'_R \left( \frac{2 M'}{M} - \frac{B'}{B} \right) + 2 f''_R - B f = 0, \label{veq-1}
\end{equation}
\begin{equation}
f_R \left( \frac{M' A'}{M A} + \frac{M'^2}{2 M^2} \right) + f'_R \left( \frac{2 M'}{M} + \frac{A'}{A} \right) - \frac{\epsilon}{2} f_X \phi'^2 - B \left[ f + \left( \frac{2}{M} -R \right) f_R - X f_X \right] = 0, \label{veq-2}
\end{equation}
\begin{equation}
f_R \left( \frac{A''}{A} + \frac{M''}{M} -  \frac{A' B'}{2 A B} + \frac{M' A'}{2 M A} - \frac{M' B'}{2 M B} - \frac{A'^2}{2 A^2} - \frac{M'^2}{2 M^2} \right) + f'_R \left( \frac{A'}{A} - \frac{B'}{B} + \frac{M'}{M} \right) + 2 f''_R  - B \left( f - R f_R \right)  = 0, \label{veq-3}
\end{equation}
\begin{equation}
\left[ M f_X \sqrt{\frac{A}{B}} \phi'  \right]' + \epsilon M \sqrt{A B} f_{\phi} = 0. \label{veq-4}
\end{equation}
The latter equation is the modified Klein-Gordon equation given in (\ref{kgeq}). Since the equation of motion (\ref{veq-2}) describing the evolution of the metric potential $B$ does not depend on its derivative, it can be explicitly solved in terms of $B$ as a function of other coordinates such that
\begin{equation}
B = \frac{f_R \left( \frac{M' A'}{M A} + \frac{M'^2}{2 M^2} \right) + f'_R \left( \frac{A'}{A} + \frac{2 M'}{M} \right) - \frac{\epsilon}{2} f_X \phi'^2}{f + \left( \frac{2}{M} -R \right) f_R - X f_X} . \label{veq-2-2}
\end{equation}
By inserting the Ricci scalar $R$ given by (\ref{r-scalar}) into the equations (\ref{veq-1}) and (\ref{veq-3}), one get
\begin{equation}
f_R \left( \frac{2 M''}{M} - \frac{M' A'}{M A} - \frac{M' B'}{M B} - \frac{M'^2}{M^2} \right) - f'_R \left( \frac{A'}{A} + \frac{B'}{B} \right) + 2 f''_R + \epsilon f_X \phi'^2 = 0, \label{veq-1-2}
\end{equation}
and
\begin{equation}
f_R \left( \frac{M''}{M} +  \frac{M' A'}{2 M A} - \frac{M' B'}{2 M B} - \frac{2 B}{M} \right) - f'_R \left( \frac{A'}{A} - \frac{B'}{B} + \frac{M'}{M} \right) - 2 f''_R  + B f = 0, \label{veq-3-2}
\end{equation}
in which the Eq.(\ref{veq-1-2}) comes from the field equations (\ref{feq1}) and (\ref{feq2}), and the Eq. (\ref{veq-3-2}) is equivalent to the field equation (\ref{feq3}).
The energy functional $E_{\mathcal{L}}$ or the Hamiltonian of the Lagrangian $\mathcal{L}$  is defined by
\begin{eqnarray}
E_{\mathcal{L}} = q'^i \frac{\partial \mathcal{L}}{\partial q'^i} -\mathcal{L}. \label{energy}
\end{eqnarray}
Now, we calculate the energy functional $E_{\mathcal{L}_f}$ for the Lagrangian density $\mathcal{L}_f$ which has the form
\begin{eqnarray}
& & E_{\mathcal{L}_f} = M \sqrt{\frac{A}{B}} \left\{ f_R \left( \frac{M' A'}{M A} +  \frac{M'^2 }{2 M^2} \right) + f'_R \left( \frac{A'}{A} + 2 \frac{M'}{M} \right)  - \frac{\epsilon}{2} f_X \phi'^2 - \frac{B}{M} \left[ M (f - X f_X) + (2 - MR) f_R \right] \right\} . \label{ef1}
\end{eqnarray}
It is explicitly seen that the energy function  $E_{\mathcal{L}_f}$ vanishes due to the field equation (\ref{veq-2}) which is obtained by varying the Lagrangian (\ref{lagr}) according to the metric variable $B$. Therefore, the solution of equation $E_{\mathcal{L}_f} = 0$ in terms of $B$ is given by (\ref{veq-2-2}).

We note that the Hessian determinant of the Lagrangian (\ref{lagr}), which is defined by $\| \frac{\partial^2 \mathcal{L}_f }{\partial q'^i \partial q'^j}  \|$, is zero. This is because of the absence of the generalized velocity $B$ in the symmetry reduced Lagrangian (\ref{lagr}). It is known that the metric variable $B$ does not contribute to the dynamics due to the symmetry reduced Lagrangian approach, but the equation of motion for $B$ has to be considered as a further constraint equation. Thus, the new Lagrangian reads $ \mathcal{L}^*_f =  \mathcal{L}^{1/2}$ with
\begin{eqnarray}
\mathcal{L}= \left[ M ( f -X f_X) + (2- M R) f_R \right] \left[ f_R \left(  M' A' + \frac{A M'^2}{2 M} \right) +  f'_R \left( M A'  + 2 A M' \right)  - \frac{\epsilon}{2} f_X  M A \phi'^2 \right],  \qquad \label{lagr2}
\end{eqnarray}
which is not explicitly dependent on $r$, so it is a canonical Lagrangian, and the quadratic form of generalized velocities $A', M', R', \phi'$ and $X'$ and thus coincides with the Hamiltonian. Therefore, we can consider $\mathcal{L}$ as the new Lagrangian with five degrees of freedom. The Hessian determinant of $\mathcal{L}$  is still vanishing, which comes from the selection of $X$ as a coordinate of the configuration space. This new Lagrangian has a property that if $f(R,\phi,X) = F(R)$, then the Hessian determinant of the Lagrangian is now non-vanishing as it is expected. Therefore it has to be supposed that $M ( f -X f_X) + (2- M R) f_R \neq 0$, which is necessary since the above definitions of $B$, the Eq. (\ref{veq-2-2}), and $\mathcal{L}$, the Eq. (\ref{lagr2}), lose significance. Furthermore, using the Eq. (\ref{energy}) we calculate the energy function for the new Lagrangian (\ref{lagr2}) and find that
\begin{eqnarray}
& & E_{\mathcal{L}} = A B  \left[ M (f - X f_X) + (2 - MR) f_R \right]^2, \label{ef2}
\end{eqnarray}
which is explicitly non-vanishing. The new Lagrangian (\ref{lagr2}) is useful to compare the Noether symmetries obtained in this study with the results of Ref. \cite{capo2007}, where they have been selected families of $f(R)$ models in which the spherical symmetry has been imposed, and searched for exact spherically symmetric solutions in $f(R)$ gravity by requiring the existence of Noether symmetries. In order to facilitate the comparison with the results of Ref. \cite{capo2007} for the power-law form $f(R,\phi, X) = f_0 R^n$ we use the Lagrangian (\ref{lagr2}) to compute the Noether symmetries.

\bigskip

%%%%%%%%%%%%%%%%%%%%%%%%%%%
%%%  Sec. IV
%%%%%%%%%%%%%%%%%%%%%%%%%%%
\section{Noether symmetry approach}

In this section, we seek for the condition in order that the Lagrangian density (\ref{lagr}) or (\ref{lagr2}) would admit any Noether symmetry which has a generator of the form  %\cite{cimall}
\begin{equation}
{\bf Y} = \xi \frac{\partial}{\partial r} + \eta^i \frac{\partial}{\partial q^i}, \label{ngs-gen}
\end{equation}
where $q^i$ are the generalized coordinates in the $d$-dimensional configuration space ${\cal Q }\equiv \{ q^i, i=1, \ldots, d \} $ of the Lagrangian, whose tangent space is ${\cal TQ }\equiv \{q^i,q'^i\}$. The components $\xi$ and $\eta^i$ of the Noether symmetry generator ${\bf Y}$ are functions of $r$ and $q^i$. The existence of a Noether symmetry implies the existence of a vector field ${\bf Y}$ given in (\ref{ngs-gen}) if the Lagrangian $ \mathcal{L}(r, q^i, q'^i )$ satisfies
\begin{equation}
{\bf Y}^{[1]} \mathcal{L} + \mathcal{L} ( D_r \xi) = D_r K\, , \label{ngs-eq}
\end{equation}
where ${\bf Y}^{[1]}$ is the first prolongation of the generator (\ref{ngs-gen}) in such a form
\begin{equation}
{\bf Y}^{[1]} = {\bf Y}  + \eta'^i \frac{\partial}{\partial q'^i},
\end{equation}
and $K(r, q^i)$ is a gauge function, $D_r$ is the total derivative operator with respect to $r$, $D_r =\partial / \partial r + q'^i \partial / \partial q^i$, and $\eta'^i$ is defined as $\eta'^i = D_r \eta^i - q'^i D_r \xi$. The significance of Noether symmetry comes from the following first integral of motion that if ${\bf Y}$ is the Noether symmetry generator corresponding to the Lagrangian $\mathcal{L}(r, q^i, q'^i)$, then the Hamiltonian or a conserved quantity associated with the generator ${\bf Y}$ is
\begin{equation}
I =- \xi E_{\mathcal{L}} + \eta^i \frac{\partial \mathcal{L}}{\partial q'^i} - K , \label{con-law}
\end{equation}
where $I$ is a constant of motion \emph{or} Noether constant.

Let us start with the Lagrangian (\ref{lagr2}), where $q^i = \{ A, M, R, \phi, X \}, i=1,\ldots, 5 $. Then the Noether symmetry condition (\ref{ngs-eq}) for this Lagrangian yields 26 partial differential equations as follows
\begin{eqnarray}
& & \xi_{,A} = 0, \quad \xi_{,M} = 0, \quad  \xi_{,R} = 0, \quad \xi_{,\phi} = 0, \quad \xi_{,X} = 0, \quad K_{,r} = 0, \nonumber \\& & f_R \eta^2_{,r} + M \left( f_{RR} \eta^3_{,r} + f_{R\phi} \eta^4_{,r} + f_{RX} \eta^5_{,r}  \right) - \frac{1}{F} K_{,A} =0, \quad f_{RX} \left( M \eta^1_{,r} + A \eta^2_{,r} \right) - \frac{1}{F} K_{,X} =0, \nonumber\\& & f_R \left( \eta^1_{,r} + \frac{A}{M} \eta^2_{,r} \right) + 2 A \left( f_{RR} \eta^3_{,r} + f_{R\phi} \eta^4_{,r} + f_{RX} \eta^5_{,r}  \right) - \frac{1}{F} K_{,M} =0, \nonumber\\& & f_{RR} \left( M \eta^1_{,r} + A \eta^2_{,r} \right) - \frac{1}{F} K_{,R} =0, \quad f_{RR} \left( M \eta^1_{,r} + A \eta^2_{,r} \right) - \epsilon M A f_X \eta^4_{,r} - \frac{1}{F} K_{,\phi} =0, \nonumber \\& & f_{RR} \left( M \eta^1_{,R} + 2 A \eta^2_{,R} \right) = 0, \quad f_{RR} \left( M \eta^1_{,\phi} + 2 A \eta^2_{,\phi} \right) + f_{R\phi} \left( M \eta^1_{,R} + 2 A \eta^2_{,R} \right) = 0, \nonumber \\& & f_{RR} \left( M \eta^1_{,X} + 2 A \eta^2_{,X} \right) + f_{RX} \left( M \eta^1_{,R} + 2 A \eta^2_{,R} \right) = 0, \quad  f_{RX} \left( M \eta^1_{,X} + 2 A \eta^2_{,X} \right) = 0, \nonumber\\& & f_{R\phi} \left( M \eta^1_{,X} + 2 A \eta^2_{,X} \right) + f_{RX} \left( M \eta^1_{,\phi} + 2 A \eta^2_{,\phi} \right) - \epsilon M A f_X \eta^4_{,X} = 0, \quad f_R \eta^2_{,A} + M \left( f_{RR} \eta^3_{,A} + f_{R\phi} \eta^4_{,A} + f_{RX} \eta^5_{,A}  \right) = 0, \nonumber
\end{eqnarray}
\begin{eqnarray}
& & f_R \left(  \eta^1_{,A} + \frac{A}{M} \eta^2_{,A} + \eta^2_{,M} - \xi_{,r} \right) + M \left( f_{RR} \eta^3_{,M} + f_{R\phi} \eta^4_{,M} + f_{RX} \eta^5_{,M}  \right) + 2 A \left( f_{RR} \eta^3_{,A} + f_{R\phi} \eta^4_{,A} + f_{RX} \eta^5_{,A}  \right) \nonumber \\& & \qquad + \frac{1}{F}\left[ f_R F_{,M} \eta^2 + \left( F f_R \right)_{,R} \eta^3 + \left( F f_R \right)_{,\phi} \eta^4 + \left( F f_R \right)_{,X} \eta^5 \right] = 0, \nonumber
\end{eqnarray}
\begin{eqnarray}
& &  \frac{f_R}{M} \eta^2_{,R} +  f_{RR} \left(  \eta^1_{,A} + \frac{2 A}{M} \eta^2_{,A} + \eta^3_{,R} - \xi_{,r} \right) +  f_{R\phi} \eta^4_{,R} + f_{RX} \eta^5_{,R}  \nonumber \\& & \qquad + \frac{1}{F} \left[ f_{RR} (F M)_{,M} \frac{\eta^2}{M} + \left( F f_{RR} \right)_{,R} \eta^3 + \left( F f_{RR} \right)_{,\phi} \eta^4 + \left( F f_{RR} \right)_{,X} \eta^5 \right] = 0, \nonumber
\end{eqnarray}
\begin{eqnarray}
& &  \frac{f_R}{M} \eta^2_{,\phi} +  f_{R\phi} \left(  \eta^1_{,A} + \frac{2 A}{M} \eta^2_{,A} + \eta^4_{,\phi} - \xi_{,r} \right) +  f_{RR} \eta^3_{,\phi} + f_{RX} \eta^5_{,\phi}  \nonumber \\& & \qquad + \frac{1}{F} \left[ f_{R\phi} (F M)_{,M} \frac{\eta^2}{M} + \left( F f_{R\phi} \right)_{,R} \eta^3 + \left( F f_{R\phi} \right)_{,\phi} \eta^4 + \left( F f_{R\phi} \right)_{,X} \eta^5 \right] = 0, \label{ngs-eq-1}
\end{eqnarray}
\begin{eqnarray}
& &  \frac{f_R}{M} \eta^2_{,X} +  f_{RX} \left(  \eta^1_{,A} + \frac{2 A}{M} \eta^2_{,A} + \eta^5_{,X} - \xi_{,r} \right) +  f_{RR} \eta^3_{,X} + f_{R\phi} \eta^4_{,X}  \nonumber \\& & \qquad + \frac{1}{F} \left[ f_{RX} (F M)_{,M} \frac{\eta^2}{M} + \left( F f_{RX} \right)_{,R} \eta^3 + \left( F f_{RX} \right)_{,\phi} \eta^4 + \left( F f_{RX} \right)_{,X} \eta^5 \right] = 0, \nonumber
\end{eqnarray}
\begin{eqnarray}
& & f_R \left[ \frac{\eta^1}{A} +  \frac{M}{F} \left( \frac{F}{M} \right)_{,M} \eta^2 + \frac{2 M}{A} \eta^1_{,M} + 2 \eta^2_{,M} - \xi_{,r} \right] + 4 M \left( f_{RR} \eta^3_{,M} + f_{R\phi} \eta^4_{,M} + f_{RX} \eta^5_{,M}  \right) \nonumber \\& & \qquad   + \frac{1}{F}\left[ \left( F f_R \right)_{,R} \eta^3 + \left( F f_R \right)_{,\phi} \eta^4 + \left( F f_R \right)_{,X} \eta^5 \right] = 0, \nonumber
\end{eqnarray}
\begin{eqnarray}
& & \frac{f_R}{2 A} \left( \eta^1_{,R} + \frac{A}{M} \eta^2_{,R} \right) + f_{RR} \left[ \frac{\eta^1}{A} +  \frac{F_{,M}}{F}  \eta^2 + \frac{M}{2 A} \eta^1_{,M} + \eta^2_{,M} + \eta^3_{,R} - \xi_{,r} \right] + f_{R\phi} \eta^4_{,R} + f_{RX} \eta^5_{,R}  \nonumber \\& & \qquad   + \frac{1}{F}\left[ \left( F f_{RR} \right)_{,R} \eta^3 + \left( F f_{RR} \right)_{,\phi} \eta^4 + \left( F f_{RR} \right)_{,X} \eta^5 \right] = 0, \nonumber
\end{eqnarray}
\begin{eqnarray}
& & \frac{f_R}{2 A} \left( \eta^1_{,\phi} + \frac{A}{M} \eta^2_{,\phi} \right) + f_{R\phi} \left[ \frac{\eta^1}{A} +  \frac{F_{,M}}{F}  \eta^2 + \frac{M}{2 A} \eta^1_{,M} + \eta^2_{,M} + \eta^4_{,\phi} - \xi_{,r} \right] + f_{RR} \eta^3_{,\phi} + f_{RX} \eta^5_{,\phi}  \nonumber \\& & \qquad   + \frac{1}{F} \left[ \left( F f_{R\phi} \right)_{,R} \eta^3 + \left( F f_{R\phi} \right)_{,\phi} \eta^4 + \left( F f_{R\phi} \right)_{,X} \eta^5 \right] = 0, \nonumber
\end{eqnarray}
\begin{eqnarray}
& & \frac{f_R}{2 A} \left( \eta^1_{,X} + \frac{A}{M} \eta^2_{,X} \right) + f_{RX} \left[ \frac{\eta^1}{A} +  \frac{F_{,M}}{F} \eta^2 + \frac{M}{2 A} \eta^1_{,M} + \eta^2_{,M} + \eta^5_{,X} - \xi_{,r} \right] + f_{RR} \eta^3_{,X} + f_{R\phi} \eta^4_{,X}  \nonumber \\& & \qquad   + \frac{1}{F} \left[ \left( F f_{RX} \right)_{,R} \eta^3 + \left( F f_{RX} \right)_{,\phi} \eta^4 + \left( F f_{RX} \right)_{,X} \eta^5 \right] = 0, \nonumber
\end{eqnarray}
\begin{eqnarray}
& & f_X \left[ \frac{\eta^1}{A} + \frac{(M F)_{,M}}{M F} \eta^2 + 2 \eta^4_{\phi} - \xi_{,r} \right] - \frac{2 \epsilon}{M A} f_{R\phi} \left( M \eta^1_{,\phi} + 2 A \eta^2_{,\phi} \right) + \nonumber \\& & \qquad \qquad  + \frac{1}{F} \left[ \left( F f_{X} \right)_{,R} \eta^3 + \left( F f_{R} \right)_{,\phi} \eta^4 + \left( F f_{X} \right)_{,X} \eta^5 \right] = 0, \nonumber
\end{eqnarray}
where $F = \left[ M ( f -X f_X) + (2- M R) f_R \right] \neq 0$. Then we solve the above system of differential equations to get the Noether symmetry ${\bf Y}= \xi \partial_r + \eta^1 \partial_A + \eta^2 \partial_M + \eta^3 \partial_R + \eta^4 \partial_{\phi} + \eta^5 \partial_X$. The above system implicitly depends on the form of the function $f(R, \phi,X)$ and so, by solving it, we also get a wide class of gravity theories related to the form of $f(R,\phi,X)$ which are compatible with spherical symmetry. We point out here that the Noether symmetries for any form of the function $f(R,\phi,X)$ are
\begin{equation}
{\bf Y}_1 = \partial_r, \qquad {\bf Y}_2 = r \partial_r +  A \partial_A, \label{ns-12}
\end{equation}
which are trivial solutions of the above system in any case. In the following we will consider some forms of $f(R,\phi,X)$ to search the Noether symmetries. To do this, we will split the study in five different types of $f(R,\phi,X)$:
\begin{enumerate}
		\item $f(R,\phi,X)=f_0 R^n$, where $f_0$ and $n$ are constants (see \ref{case1}).
		\item $f(R,\phi,X)= f_0 R + f_1 X^q - V(\phi)$, where $f_0,f_1$ and $q$ are constants and $V(\phi)$ is a potential (see \ref{case2}).
		\item $f(R,\phi,X)= f_0 \phi^m R^n + f_1 X^q - V(\phi)$, where $f_0,f_1,m,n$ and $q$ are constants and $V(\phi)$ is a potential (see \ref{case3}).
		\item $f(R,\phi,X)= f_0 R^n + f_1 \, \phi^m\, X^q$, where $f_0,f_1,m,n$ and $q$ are constants (see \ref{case4}).
		\item $f(R,\phi,X)=  U(\phi, X) \, R$, where $U(\phi,X)$ is an arbitrary function which depends on the scalar field $\phi$ and the kinetic term $X$ (see \ref{case5}).
	\end{enumerate}
In each case, some subcases where some specific parameters will be also studied. The specific cases listed above represent different classes of modified theories of gravity that can be constructed with the curvature scalar $X$, a scalar field $\phi$ and a kinetic term $X$. The first model has been widely considered in the literature and represents a power-law $f(R)$ gravity~\cite{Sotiriou:2008rp,Nojiri:2010wj,DeFelice:2010aj}.
These models provide a good fitting for galactic rotation curves~\cite{Martins:2007uf}. The second model represents a generalisation of minimally coupled models (quintessence models) where the parameter $q$ gives the opportunity to have power-law kinetic terms in the action. Similar models have been studied in \cite{Capozziello:2002rd,Amendola:1999er}. The third model represents a class of non-minimally couplings between a scalar field and the Ricci scalar. The case $m=2$ and $n=1$ has been widely studied in the literature, and in cosmology, it can reproduce a late-time accelerating scenarios with the possibility of crossing the phantom barrier~\cite{Perrotta:1999am,Copeland:2006wr}. Other studies have considered other power-law parameters, see for example \cite{Sami:2012uh,Gannouji:2006jm,Elizalde:2008yf}. The fourth model is another family of coupling models where now the kinetic term is coupled with the scalar field. This model is a particular case of the $k$-essence models, which can describe late-time accelerating behaviour of the Universe and can describe inflation in a good agreement with cosmological observations~\cite{ArmendarizPicon:2000ah,Deffayet:2011gz,ArmendarizPicon:1999rj}. Moreover, these models also can achieve a unified model for dark matter and dark energy avoiding the problems of the generalized Chaplygin gas models, which are due to a non-negligible sound speed in these models~\cite{Scherrer:2004au}. Finally, the last model represents a Brans-Dicke theory, see~\cite{Brans:1961sx,Amendola:1999er} for more details about this theory. All these models have been widely studied in the literature in the context of cosmology but due to the difficulty of finding exact solutions, there are not so many studies concerning non-trivial spherically symmetric spacetimes. In this paper, we will find exact solutions to the models since they have interesting physics properties.

\bigskip

\subsection{\bf Case (i): $f(R,\phi,X) = f_0 R^n$.\label{case1}}
This power-law form of $f(R,\phi,X)$  gives the well-known $f(R)$ theory of gravity, and Noether symmetries have been investigated in Ref. \cite{capo2007}. As an example to see how our approach works, we revisit this  form of $f(R,\phi,X)$. Furthermore, the trace Eq.(\ref{trace}) in this case becomes
\begin{equation}
3 n \square \left( R^{n-1} \right) + (n-2) R^n = 0,
\end{equation}
which gives $R= 0$ for $n=1$, $\square R = 0$ for $n=2$, etc. We find from the system (\ref{ngs-eq}) that the components of the Noether generator ${\bf Y}$ for this case are
\begin{equation}
\xi = c_1 r +c_2, \quad \eta^1 = \left[ c_1 + c_3 (2n -3) \right] A, \quad \eta^2 = c_3 M, \quad \eta^3 = -c_3 R, \quad \eta^4 = 0 = \eta^5, \quad K= c_4,
\end{equation}
which yields that ${\bf Y}_1 , \, {\bf Y}_2$ and
\begin{eqnarray} \label{ngsv-1}
& & {\bf Y}_3 = (2n-3) A \partial_A + M \partial_M - R \partial_R, \quad n \neq 0,1,2
\end{eqnarray}
are Noether symmetries. This explicitly represents that there exist extra two Noether symmetries ${\bf Y}_1$ and ${\bf Y}_2$ in addition to the known one ${\bf Y}_3$ found in \cite{capo2007}. The first integrals of the above Noether symmetries are
\begin{eqnarray}
& & I_1 = -  E_{\mathcal{L}}, \qquad I_2 =  I_1 r + f_0^2 n M A R^{2 (n-1)} \left[ 2n + (1-n) M R \right] \left[ (n-1) \frac{R'}{R} + \frac{M'}{M} \right], \label{fint-1-1} \\& &  I_3 =  f_0^2 n M A R^{2 (n-1)} \left[ 2n + (1-n) M R \right] \left[ (n-1)(2n -1) \frac{R'}{R} + (2- n) \frac{A'}{A} \right] , \label{fint-1-2}
\end{eqnarray}
where $I_1$ is non-vanishing due to the $E_{\mathcal{L}} \neq 0$.
Then, arranging the above first integrals one gets
\begin{eqnarray}
& & (n-1) \frac{R'}{R} \left( \frac{A'}{A} + \frac{2 M'}{M} \right) + \frac{M'}{M} \left( \frac{A'}{A} + \frac{M'}{2 M} \right) = -\frac{ I_1 R^{2 (1-n)} }{ n f_0^2 M A \left[ 2n + (1-n) M R \right] }, \nonumber \\& & (n-1) \frac{R'}{R} + \frac{M'}{M}  = \frac{ (I_2 - I_1 r)  R^{2 (1-n)} }{ n f_0^2 M A \left[ 2n + (1-n) M R \right] }, \label{fint-1-1-1} \\& &  (2- n) \frac{A'}{A} + (n-1)(2n -1) \frac{R'}{R}  = \frac{ I_3 R^{2 (1-n)}  }{ n f_0^2 M A  \left[ 2n + (1-n) M R \right]} , \nonumber
\end{eqnarray}
where $n \neq 1, 2$. Solving the third equation of (\ref{fint-1-1-1}) in terms of $A$, one finds
\begin{equation}
A = R^{\frac{(n-1)(2n-1)}{n-2}} \left[ A_0 + \frac{I_3}{ f_0^2 n (2-n)} \int{ \frac{ R^{\frac{(n-1)(4n-5)}{2-n}} dr }{M [2n + (1-n) M R ] } }  \right], \label{i-A}
\end{equation}
which has same form obtained in \cite{capo2007}. Due to the previously obtained relation (\ref{veq-2-2}) of the metric function $B$, it takes the form for this case:
\begin{equation}
B =  \frac{ n M }{ \left[ 2n + (1-n) M R  \right] } \left[ (n-1) \frac{R'}{R} \left( \frac{A'}{A} + \frac{2 M'}{M} \right) + \frac{M'}{M} \left( \frac{A'}{A} + \frac{M'}{2 M} \right) \right]. \label{i-B}
\end{equation}
We observe here that one can find the metric functions $A$ and $B$ if the functions $M(r)$ and $R(r)$ are known one way or another.
One can give so much examples of the exact solutions for the field equations using the above relations (\ref{i-A}), (\ref{fint-1-1-1}) and (\ref{i-B}). Here the Eqs. (\ref{fint-1-1-1}) are constraint equations to be satisfied.
If one chooses $M(r)=r^2$ and $R=R_0 r^{p}$, one gets hypergeometric functions for both $A(r)$ and $B(r)$. There are some specific cases for $p$ where one can get analytical solutions without those hypergeometric functions. The easiest case is as it was chosen in Ref. \cite{capo2007}, where one takes $n=5/4,p =-2 ,R_0=-5$ which gives $R(r)= - \frac{5}{r^2}$. The minus sign in $R$ is due to the signature of the metric. For this case, the metric coefficients $A$ and $B$ from the relations (\ref{i-B}) and (\ref{i-A}) are obtained as
\begin{equation}
A(r)= \frac{1}{\sqrt{5}} \left( k_1 r + k_2 \right), \qquad B(r)= \frac{1}{ 2\left(1 + \frac{k_2}{k_1 r } \right) },
\end{equation}
where $k_1 = A_0, k_2 = 32 I_2 /(225 f_0^2), I_1 = - I_2 k_1 /k_2$ and $I_3  = -I_2/2$ which comes from the constraint equations (\ref{fint-1-1-1}).  The latter solution was already found in \cite{capo2007}. It should be noted that the above metric is non asymptotically flat, has a horizon at $r=-k_2/k_1$ and it was ruled out by Solar system tests~\cite{Clifton:2005aj}. Another new solution can be found by taking  $M(r)=r^2, R=R_0 r^{p}$ and $p=(n-2)/(4 n^2-10 n+7)$, which gives the following metric coefficient
\begin{eqnarray}
A(r)&=& r^{\frac{(n-1) (2 n-1)}{4 n^2 -10 n+7}}  \left[  A_0 R_0^{\frac{2 n^2-3 n+1}{n-2}}  + \frac{I_3 (n-1) \left(4 n^2-10 n+7 \right) R_0^{3-2 n} }{4 f_0^2 n^3 (n-2) \left(8 n^2-19 n+12\right)} \log \left(1-\frac{2 n r^{\frac{-8 n^2+19 n-12}{4 n^2-10 n+7}}}{(n-1) R_0}\right) \right] \nonumber\\
&& \quad + \frac{ I_3 \left(4 n^2-10 n+7\right) R_0^{2-2 n} }{2 f_0^2 n^2 (n-2) \left(8 n^2-19 n+12\right)}r^{\frac{-6 n^2+16 n-11}{4 n^2-10 n+7}}\,.  \label{AA1}
\end{eqnarray}
The expression for $B(r)$ is involved but can be directly found by using Eq.~\eqref{i-B}. Another analytical solution can be found from \eqref{i-A} by taking $M(r)=r^2, R=R_0 r^{p}$ and $p=(2-n)/(4 n^2-9 n+5)$. This solution reads
\begin{eqnarray}
A(r)&=&r^{\frac{1-2 n}{4 n-5}} R_0^{\frac{2 n^2-3 n+1}{n-2}} \left[ A_0+\frac{I_3 (n-1) (4 n-5) R_0^{\frac{-4 n^2+9 n-5}{n-2}} \log \left((n-1) R_0 r^{\frac{8 n^2-19 n+12}{4 n^2-9 n+5}}-2 n\right)}{2 f_0^2 (n-2) n^2 (n (8 n-19)+12)} \right] \nonumber\\
&&-\frac{I_3 r^{\frac{1-2 n}{4 n-5}} R_0^{2-2 n} \log (r)}{2 f_0^2 (n-2) n^2}\,,\label{AA2}
\end{eqnarray}
with $B(r)$ being also too involved to write it here but it can be easily found with Eq.~\eqref{i-B}.

It is possible to give some other examples to produce new solutions from the generic statements for $A$ and $B$. If we take $n= 1/2, M(r) = r^q$ and $R(r) = R_0 r^{-q}$, then it follows from \eqref{r-scalar}, \eqref{i-A} and \eqref{i-B} that $q=2/3$ and $R_0 = 1$, which gives
\begin{eqnarray}
& & A(r) = A_0 \left( 1 - \frac{2 k}{r^{2/3}} \right), \qquad B(r) = \frac{2}{21 r^{4/3} \left( 1 - \frac{2 k}{r^{2/3}} \right)},\label{47}
\end{eqnarray}
where $k = 2 I_3 / (3 A_0 f_0^2)$. This solution has an event horizon at $r = ( 2 k)^{3/2}$ and it is asymptotically flat. Furthermore, taking $M(r) = r^q$ and $R(r) = R_0 r^{-q}$, the equations \eqref{r-scalar}, \eqref{i-A} and \eqref{i-B} yield $q= 2 / 43$ and $R_0 = 1006 / 321$ for $n=3$, and $q= 4 / 167$ and $R_0 = 72 / 23$ for $n=4$, which gives rise to the solutions $A$ and $B$
\begin{eqnarray}
& & A(r) = \frac{A_0 R_0^{10}}{r^{20/43}}  \left[ 1 + \frac{43 k R_0^{14}}{13 (R_0 -3)} r^{13/43} \right], \quad B(r) = - \frac{r^{-84/43}}{86  \left( R_0 -3 + \frac{43}{13} k R_0^{14} r^{13/43} \right)}, \quad k = \frac{I_3}{ 6 A_0 f_0^2},\label{48}
\end{eqnarray}
for $n=3$, and
\begin{eqnarray}
& & A(r) = \frac{A_0 R_0^{21/2}}{r^{42/167}}  \left[ 1 + \frac{167 k R_0^{33}}{ (3 R_0 -8)} r^{31/167} \right], \quad B(r) = - \frac{ 30752 \, r^{-330/167}}{ 27889 \left( 93 R_0 - 248 + 167 k R_0^{33} r^{31/167} \right)}, \quad k = \frac{I_3}{ 8 A_0 f_0^2}, \label{49}
\end{eqnarray}
for $n=4$. These two solutions are also asymptotically flat. To the best of our knowledge, the five solutions \eqref{AA1}--\eqref{49} are new spherically symmetric solutions in power-law $f(R)$ gravity. It is important to remark that in the literature, many authors have found solutions in $f(R)$ only considering $R=\textrm{constant}$, which indeed is a trivial case since all the higher order terms considered in $f(R)$ disappears. The above new solutions could be interesting since they contain some logarithmic terms that could be related to dark matter~\cite{Li:2012zx}.

\bigskip

For $n=1$ (the GR case), i.e., $f(R,\phi,X) = f_0 R$, the reduced Lagrangian and the constraint (\ref{veq-2-2}) for $B$ become
\begin{equation}
\mathcal{L}_{GR} = A M' \left( \frac{2 A'}{A} + \frac{M'}{M} \right), \qquad B_{GR} = M' \left( \frac{2 A'}{A} + \frac{M'}{M} \right) ,
\end{equation}
which yields that the energy function is $E_{\mathcal{L}_{GR}} = A \, B_{GR}$. Then we find from the Noether symmetry condition (\ref{ngs-eq}) for the above Lagrangian the following Noether symmetries
\begin{eqnarray}
& & {\bf Y}_1 = \partial_r, \quad {\bf Y}_2 = r \partial_r + A \partial_A, \quad {\bf Y}_3 = A \partial_A - M \partial_M, \quad {\bf Y}_4 = \frac{1}{\sqrt{M}} \partial_A,  \\& & {\bf Y}_5 = \frac{A}{\sqrt{M}} \partial_A - 2 \sqrt{M} \partial_M, \quad {\bf Y}_6 = r^2 \partial_r + 2 r M \partial_M \quad {\rm with} \,\, K = 4 M A, \\& & {\bf Y}_7 = \frac{r A}{\sqrt{M}} \partial_A - 2 r \sqrt{M} \partial_M \quad {\rm with} \,\, K = -4 A \sqrt{M}, \quad {\bf Y}_8 = \frac{r}{\sqrt{M}} \partial_A  \quad {\rm with} \,\, K = 4 \sqrt{M},
\end{eqnarray}
which gives rise to the following first integrals
\begin{eqnarray}
& & I_1 = -E_{\mathcal{L}_{GR}}, \quad I_2 = I_1 r  + 2 A M', \quad I_3 = -2 M A', \quad I_4 = \frac{2 M'}{\sqrt{M}}, \quad I_5 = - 2 A \sqrt{M} \left( 2\frac{A'}{A} + \frac{M'}{M} \right), \\& & I_6 = I_1 r^2 + 4r M A \left( \frac{A'}{A} + \frac{M'}{M}  - \frac{1}{r} \right), \quad I_7 = - 2 r A \sqrt{M} \left( \frac{2 A'}{A} + \frac{M'}{M} - \frac{2}{r} \right), \quad I_8 = I_4 r - 4 \sqrt{M}.
\end{eqnarray}
Here we point out that the Noether symmetry ${\bf Y}_3$ for GR Lagrangian has only been obtained in \cite{capo2007}. Using the above first integrals, the functions $A$ and $M$ together have the Schwarzschild form with some constraints as follows
\begin{eqnarray}
& & A = \frac{I_7 - I_5 r}{I_4 r -I_8}, \qquad B= -\frac{I_1}{A}, \qquad M= \frac{1}{16} \left( I_4 r - I_8 \right)^2,  \\& & I_5 = \frac{ 4 I_1}{I_4}, \quad I_6 = \frac{I_2}{I_1} (I_2 - 2 I_3), \quad I_7 = \frac{4 I_2}{I_4}, \quad I_8 = \frac{I_4}{I_1} (I_2 - 2 I_3).
\end{eqnarray}
Thus the standard form of Schwarzschild solution is covered for $I_1 =-1, I_4 = -4, I_5 = 1, I_6 = 0, I_8 = 0$ and $I_2 = 2 I_3 = -I_7 = 8 m$, where $m$ is the Schwarzschild mass, which means that the Noether symmetry relates the first integrals $I_2, I_3$ and $I_7$ with the Schwarzschild radius or the mass of the gravitating system.  For the classical Schwarzschild solution, the event horizon at $r = 2m$ corresponds to a singularity of the Schwarzschild coordinates at which $g_{00} = 0$, i.e. $A(r) = 0$, and $g_{11}= B(r)$ tends to infinity. It is known that a horizon is a null-hypersurface, and one can say that $r={\rm constant}$ is the null-hypersurface at $A(r) = 0$ which yields all the possible horizons.

\bigskip

For $n=2$, the Noether symmetries are ${\bf Y}_1, \, {\bf Y}_2 $ given in (\ref{ns-12}) and
\begin{equation}
{\bf Y}_3 = A \partial_A + M \partial_M - R \partial_R,
\end{equation}
with the first integrals $I_1 = - E_{\mathcal{L}}$ and
\begin{equation}
I_2 = I_1 r + 2 f_0^2 M A R^2 ( 4 - M R) \left( \frac{R'}{R} + \frac{M'}{M} \right), \qquad I_3 = 6 f_0^2 M A ( 4 - M R) R R',  \label{fint-2}
\end{equation}
in which the latter first integral yields
\begin{equation}
A = \frac{I_3}{ 6 f_0^2 M (4 - M R) R R'} \, , \label{i-A-2}
\end{equation}
with $I_3 \neq 0$. Then it follows from the first integrals (\ref{fint-2}) that
\begin{equation}
\frac{R'}{R} + \frac{I_3}{(3 I_1 r + \alpha)} \frac{M'}{M} = 0,
\end{equation}
which has a solution
\begin{equation}
R = R_0 \left( 1 + \frac{ q}{ r}  \right)^{\frac{2 I_3}{\alpha}}, \label{i-R-2}
\end{equation}
for $M(r) = r^2$, where $R_0$ is an integration constant, $\alpha=I_3 - 3 I_2$ and $q= \frac{ \alpha}{3 I_1}$. Thus, using (\ref{i-R-2}) in (\ref{i-A-2}) and $I_1 = -f_0^2 A B R^2 (4 - M R)^2$, we find
\begin{equation}
A = A_0 \left( 1 + \frac{q}{r} \right)^{1- \frac{4 I_3}{ \alpha} } \left[  R_0 r^2 \left( 1 + \frac{q}{r} \right)^{\frac{2 I_3}{ \alpha} }  -4   \right]^{-1}, \quad B = \frac{4}{ \left( 1 + \frac{q}{r} \right) } \left[  4- R_0 r^2 \left( 1 + \frac{q}{r} \right)^{\frac{2 I_3}{ \alpha} }  \right]^{-1}
\end{equation}
where $A_0 = \frac{I_1}{4 f_0^2 R_0^2}$. So we have obtained a complete solution of the quadratic gravity ($n=2$) when $M (r)= r^2$. As far as we know, this solution is a new exact solution for the quadratic gravity. This solution is asymptotically flat if $0<I_3/\alpha<1/4$ and can describe a black hole since its horizons are at $r=-q$ and when $r^2R_0(q/r+1)^{2I_3/\alpha}=4$, which depends on the exponent $I_3/\alpha$. Thus, the above solution is a new black hole solution in quadratic gravity given by $f(R)=f_0R^2$.

\subsection{{\bf Case (ii):} $f(R,\phi,X)= f_0 R + f_1 X^q - V(\phi)$.\label{case2}}

In this case, the existence of a non-trivial Noether symmetry selects the form of potential function $V(\phi)$ of the theory, which means that it is possible to find out exact solutions for a given theory with the selected potential. Here the function $F$ takes the form $F = 2 f_0 + f_1 ( 1 - q) M X^q - M V(\phi)$. Further, the field Eq.~(\ref{fieldeq2}) and trace Eq.(\ref{trace}) for this case yield
\begin{equation}
R = \frac{2 V(\phi)}{f_0} + (q-2) \frac{f_1}{f_0} X^q, \qquad \nabla_{\mu} \left( X^{q-1} \nabla^{\mu} \phi \right) - \frac{\epsilon}{q f_1} V_{\phi} = 0.
\end{equation}
For the potential $V(\phi) = V_0 \left( \phi + V_1 \right)^{\frac{2 q}{1-q}}$, we find the Noether symmetries ${\bf Y}_1, \, {\bf Y}_2$ given in (\ref{ns-12}) and
\begin{eqnarray}
& & {\bf Y}_3 = A \partial_A - M \partial_M + \frac{ (1-q)}{2q} (\phi + V_1) \partial_{\phi} + \frac{X}{q} \partial_X,  \label{ns-iii-1}
\end{eqnarray}
with $q \neq 1$. Further, for the potential $V(\phi) = V_0 \left( \phi + V_1 \right)^{\frac{4 q}{2-q}}$, it follows that the Noether symmetries are ${\bf Y}_1, \, {\bf Y}_2$, and
\begin{eqnarray}
& & {\bf Y}_3 = - \frac{A}{2 \sqrt{M}} \partial_A + \sqrt{M} \partial_M + \frac{(q -2)}{4 q \sqrt{M}} (\phi + V_1) \partial_{\phi} - \frac{X}{q \sqrt{M}} \partial_X, \label{ns-iii-2}
\end{eqnarray}
where $q \neq 1, 2$.  For this case, the first integrals of Noether symmetries ${\bf Y}_1$ and ${\bf Y}_2$ are
\begin{equation}
I_1 = - E_{\mathcal{L} }, \qquad I_2 = I_1 r + f_0 A F M',  \label{ns-fint-12}
\end{equation}
which are common for all possible subcases of this case, and give
\begin{equation}
A= \frac{ I_2 - I_1 r}{f_0 F M'}, \qquad B = \frac{f_0 I_1 M'}{ (I_1 r -I_2) F}. \label{sc-iii-1-1}
\end{equation}
The above relations require that
\begin{equation}
B = - \frac{ I_1 f_0^2 M'^2 }{ (I_1 r -I_2)^2 } A. \label{sc-iii-1-1b}
\end{equation}
The first integrals for (\ref{ns-iii-1}) and (\ref{ns-iii-2}) are , respectively
\begin{equation}
I_3 = - M A F \left[ f_0 \frac{A'}{A} - \frac{1}{2} f_1 \epsilon (q-1) X^{q-1} (\phi + V_1) \phi'  \right], \label{sc-iii-1-2a}
\end{equation}
and
\begin{equation}
I_3 =  \sqrt{M} A F \left[ f_0 \left( \frac{A'}{A} + \frac{M'}{2 M} \right)  - \frac{1}{4} f_1 \epsilon (q-2) X^{q-1} (\phi + V_1) \phi'  \right] \, .  \label{sc-iii-1-2b}
\end{equation}
Using the relations given in (\ref{sc-iii-1-1}), the first integral (\ref{sc-iii-1-2a}) can be written in the form
\begin{eqnarray}
& & A' - \frac{I_3 M'}{ (I_1 r -I_2) M} A = \frac{ f_1 (q-1) \epsilon^q}{2^q f_0^{2 q-1} I_1^{q-1}} \left[ \frac{ I_1 r -I_2}{M'} \right]^{2(q-1)} (\phi + V_1) \phi'^{2q -1} A^{ 2-q},
\end{eqnarray}
where $\epsilon^q = 1$ for even $q$, and $\epsilon^q = \epsilon$ for odd $q$. This is the Bernoulli differential equation for $A$, and it has the solution
\begin{eqnarray}
& & A^{q-1} = \frac{1}{\mu_1} \left[ A_0 + \frac{f_1 \epsilon^q (q-1)^2}{2^q f_0^{2q -1} I_1^{q-1}} \int{ \mu_1 \left( \frac{I_1 r -I_2}{M'} \right)^{2 (q-1)} (\phi + V_1) \phi'^{(2q -1)} dr} \right],
\end{eqnarray}
where $\mu_1$  is the integration factor given by $\mu_1 = \exp \left[ (1-q) I_3 \int{ M' dr / (I_1 r - I_2)M} \right]$, and $A_0$ is an integration constant. Furthermore, together with (\ref{sc-iii-1-1}), the first integral (\ref{sc-iii-1-2b}) takes the form
\begin{eqnarray}
& & A' + \left[ \frac{I_3 M'}{ (I_1 r -I_2) \sqrt{M}} + \frac{M'}{2 M} \right] A = \frac{ f_1 (q-2) \epsilon^q}{2^{q+1} f_0^{2 q-1} I_1^{q-1}} \left( \frac{ I_1 r -I_2}{M'} \right)^{2(q-1)} (\phi + V_1) \phi'^{(2q -1)} A^{ 2-q},
\end{eqnarray}
which is also a Bernoulli differential equation and has the following solution
\begin{eqnarray}
& & A^{q-1} = \frac{1}{\mu_2} \left[ A_1 + \frac{f_1 \epsilon^q (q-1)(q-2) }{2^{q+1} f_0^{2q -1} I_1^{q-1}} \int{ \mu_2 \left( \frac{I_1 r -I_2}{M'} \right)^{2(q-1)} (\phi + V_1) \phi'^{(2q -1)} dr} \right],
\end{eqnarray}
where $\mu_2$  is the integration factor given by $\mu_2 = M^{(q-1)/2} \exp \left[ (q-1) I_3 \int{ M' dr / (I_1 r - I_2) \sqrt{M}} \right]$, and $A_1$ is an integration constant.
Now, we search the Noether symmetries in the following relevant subcases for $q =1,2$.

\subsubsection{{\bf Subcase (ii-a):} $q= 1$.} For this subcase, the field Eq.(\ref{fieldeq2}) and trace Eq.(\ref{trace}) for this subcase give
\begin{equation}
R = \frac{2 V(\phi)}{f_0} - \frac{f_1}{f_0} X, \qquad \square  \phi = \frac{\epsilon}{ f_1} V_{\phi}.
\end{equation}
We find that there exists Noether symmetries for the potential $V= V_0 \left( \phi + V_1 \right)^4$ such that ${\bf Y}_1, \, {\bf Y}_2$ and
\begin{eqnarray}
& & {\bf Y}_3 = - \frac{A}{2 \sqrt{M}} \partial_A + \sqrt{M} \partial_M - \frac{ ( \phi + V_1)}{4 \sqrt{M}} \partial_{\phi}, \label{ns-iii-1-1}
\end{eqnarray}
Then the first integrals of this subcase are given by (\ref{ns-fint-12}) and
\begin{eqnarray}
& & I_3 = \frac{ I_2 - I_1 r}{ 2 \sqrt{M}} + f_0 F \sqrt{M} A',
\end{eqnarray}
where $F= - M (V_0 \phi + V_1)^4 + 2 f_0$. Thus, these Noether integrals give rise to the same relation with (\ref{sc-iii-1-1}), and
\begin{eqnarray}
\frac{F'}{F} + \frac{M''}{M'} - \frac{M'}{2 M} - \frac{1}{I_1 r -I_2} \left( I_1 + I_3 \frac{M'}{\sqrt{M}} \right) = 0, \quad M' \neq 0.   \label{sc-iii-1-2}
\end{eqnarray}
Therefore, for $M= r^2$, we find that
\begin{eqnarray}
& & A= - \frac{ 1}{2 f_0 F_0 r} \left( I_1 r - I_2 \right)^{I_{13}}, \qquad B = \frac{2 f_0 I_1 r}{ F_0} \left( I_1 r - I_2 \right)^{-2+I_{13}}, \\& & \phi = \frac{1}{V_0 \sqrt{r}} \left[ 2 f_0 - F_0 \left( I_1 r - I_2 \right)^{1-I_{13} }  \right]^{\frac{1}{4}}  -V_1 ,
\end{eqnarray}
where $F_0$ is an integration constant and $I_{13}=- 2 I_3/I_1$. If $I_{13}<1$, the above metric is asymptotically flat and has an horizon at $r=I_2/I_1$ with $I_{13}\neq 0$. Hence, this solution also represents a black hole solution which to the best of our knowledge is a new solution in this specific scalar field theory.

\subsubsection{{\bf Subcase (ii-b):} $q= 2$.} In this subcase, the Eqs.(\ref{fieldeq2}) and (\ref{trace}) imply that
\begin{equation}
R = -\frac{2}{f_0} V(\phi), \qquad \nabla_{\mu} \left( X \nabla^{\mu} \phi \right) - \frac{\epsilon}{2 f_1} V_{\phi} = 0.
\end{equation}
For this subcase we find the Noether symmetries for the potentials $V= V_0 \left( \phi + V_1 \right)^{-4}$ and $V= V_1 e^{ - V_0 \phi}$, respectively,
\begin{eqnarray}
& & {\bf Y}_3 = A \partial_A - M \partial_M - \frac{ (\phi + V_1) }{4} \partial_{\phi} + \frac{X}{2} \partial_X, \label{ns-iii-2-1}
\end{eqnarray}
and
\begin{eqnarray}
& & {\bf Y}_3 = - \frac{A}{2 \sqrt{M}} \partial_A + \sqrt{M} \partial_M + \frac{1}{V_0 \sqrt{M}} \partial_{\phi} - \frac{X}{2 \sqrt{M}} \partial_X,
\end{eqnarray}
in addition to ${\bf Y}_1, \, {\bf Y}_2$. Here the Noether symmetry (\ref{ns-iii-2-1}) can be obtained by taking $q=2$  in (\ref{ns-iii-1}). Corresponding first integrals with the potential $V= V_0 \left( \phi + V_1 \right)^{-4}$ are given by (\ref{ns-fint-12}), and
\begin{equation}
I_3 =  - M F \left[ f_0 A' - \frac{ f_1 (I_1 r - I_2)^2 }{ 4 I_1 f_0^2 M'^2} (\phi + V_1) \phi'^3 \right].
\end{equation}
Then, using the first integrals of this subcase, we obtain the same form of $A$ and $B$ with (\ref{sc-iii-1-1}) and also the following constraint equation:
\begin{eqnarray}
& &  A' -  \frac{I_3 M' }{(I_2 - I_1 r) M} A  = \frac{f_1 (I_2 - I_1 r)^2 (\phi + V_1) \phi'^3}{ 4 I_1 f_0^3 M'^2} ,  \label{sc-iii-2}
\end{eqnarray}
where $M' \neq 0$. For $M =r^2$, this equation has the following solution
\begin{equation}
A(r) = \left[ \frac{I_1 r - I_2}{r} \right]^{\frac{2 I_3}{I_2}} \left\{ A_0 + \frac{f_1}{16 I_1 f_0^3} \int{ \left[ \frac{I_1 r - I_2}{r} \right]^{2 (1 -\frac{I_3}{I_2})} (\phi + V_1) \phi'^3 dr} \right\},
\end{equation}
where $A_0$ is a constant of integration. Then we find the metric function $B$ through (\ref{sc-iii-1-1b}) as
\begin{equation}
B(r) = - 4 I_1 f_0^2 \left[ \frac{I_1 r - I_2}{r} \right]^{2 (\frac{I_3}{I_2} -1)} \left\{ A_0 + \frac{f_1}{16 I_1 f_0^3} \int{ \left[ \frac{I_1 r - I_2}{r} \right]^{2 (1 -\frac{I_3}{I_2})} (\phi + V_1) \phi'^3 dr} \right\}.
\end{equation}

For the potential $V= V_1 e^{ - V_0 \phi}$, the first integrals $I_1$ and $I_2$ together with $E_{\mathcal{L}} = A B F^2$ gives the same relations obtained in (\ref{sc-iii-1-1}) and (\ref{sc-iii-1-1b}), where $F = 2 f_0 - f_1 M X^2 - f_2 V_1 M e^{-V_0 \phi}$. The first integral of ${\bf Y}_3$ becomes
\begin{equation}
I_3 = \frac{ I_2 - I_1 r}{ 2 \sqrt{M}} + \sqrt{M} A F \left[ f_0 \frac{A'}{A} + \frac{f_1}{V_0 B} \phi'^2 \right],
\end{equation}
which yields
\begin{eqnarray}
& &  A' -  \left[ \frac{I_3 M' }{(I_2 - I_1 r) \sqrt{M}} + \frac{M'}{M} \right]  A  = \frac{f_1 (I_2 - I_1 r)^2 \phi'^3}{ I_1 V_0 f_0^3 M'^2}.  \label{sc-iii-3}
\end{eqnarray}
If $M = r^2$, then the Eq. (\ref{sc-iii-3}) gives
\begin{equation}
A(r) =  \frac{ \left( I_1 r - I_2 \right)^{\frac{-2 I_3}{I_1}}}{r^2}  \left\{ A_1 + \frac{f_1}{4 I_1 V_0 f_0^3} \int{  \left( I_1 r - I_2 \right)^{2 (1 +\frac{I_3}{I_1})} \phi'^3 dr} \right\},\label{AAA3}
\end{equation}
where $A_1$ is an integration constant. Thus it follows from Eq. (\ref{sc-iii-1-1b}) for $B$ that
\begin{equation}
B(r) =  - 4 I_1 f_0^2  \left( I_1 r - I_2 \right)^{-2 (1 + \frac{I_3}{I_1} )}  \left\{ A_1 + \frac{f_1}{4 I_1 V_0 f_0^3} \int{  \left( I_1 r - I_2 \right)^{2 (1 +\frac{I_3}{I_1})} \phi'^3 dr} \right\}.\label{BBB3}
\end{equation}
Then, if one specifies the scalar field $\phi$, one can find some exact solutions. As an example, let us take the following scalar field
\begin{eqnarray}
\phi(r)= \frac{3 C_1 }{I_1-2 I_3} (I_1 r - I_2)^{\frac{I_1-2 I_3}{3 I_1}}\,,
\end{eqnarray}
where $C_1$ is a constant and $I_3 \neq I_1 /2$. Using this scalar field, the integrands in \eqref{AAA3} and \eqref{BBB3} become $C_1^3$ and therefore, it is easy to get the following solution for the metric coefficients
\begin{eqnarray}
A(r)=\frac{(I_1 r-I_2)^{-\frac{2 I_3}{I_1}} }{r^2} \left( A_1 + \frac{ A_0\, r}{I_1} \right)\,,\quad B(r)=-4 f_0^2 I_1 (I_1 r-I_2)^{-2 \left(\frac{I_3}{I_1} + 1 \right)} \left( A_1 + \frac{A_0 r}{I_1} \right)\, ,  \label{AB-sol1}
\end{eqnarray}
where $A_0 = f_1 C_1^3 /(4 f_0^3 V_0)$. It is clear that for $I_1=-2I_3$, one gets for that the metric coefficients from \eqref{AB-sol1} are 
\begin{eqnarray}
A(r)&=& A_0 + \frac{\left( A_0 I_2 - 4 A_1 I_3^2 \right) }{ 2 I_3 r} - \frac{A_1 I_2}{r^2}\,,\quad B(r)= 4 f_0^2 \left( \frac{  A_0 r - 2 A_1 I_3 }{I_2 + 2 I_3 r}  \right) \,,  \label{AB-sol2}
\end{eqnarray}
which behaves similarly as a Schwarzschild metric and additional contribution. This metric is asymptotically flat and has two horizons at the surfaces $r_1 = 2 A_1 I_3 / A_0$ and $r_2 = - I_2 /(2 I_3)$, i.e $A(r) = 0$ at these surfaces. At the surface $r= r_1$, the metric coefficient $B(r)$ vanishes, and it is infinite at the surface $r = r_2$. Therefore, the horizon at $r=r_2$ has same behaviour with the classical Schwarzschild solution. For the specific case $A_0=1, A_1 = 0, I_2 = -4 m I_3$ and $I_3= 2 f_0^2$, one gets the standard Schwarzschild solution $A(r)=1-\frac{2m}{r}$ and $B(r)=1/A(r)$ with $m= - I_2 / (4 I_3)$. Thus, the Eqs. \eqref{AB-sol2} can be understood as a generalization of the Schwarzschild solution found in General Relativity.

\subsection{{\bf Case (iii):} $f(R,\phi,X)= f_0 \phi^m R^n + f_1 X^q - V(\phi)$.\label{case3}}
For this case, the field Eq.(\ref{fieldeq2}) and trace Eq.(\ref{trace}) take the following form
\begin{eqnarray}
& & \square \left( \phi^m R^{n-1} \right) + \frac{(n-2)}{3 n} \phi^m R^n + \frac{2}{3 n f_0} \left[ \left( \frac{q}{2} -1 \right) X^q + V(\phi) \right] = 0,  \\ & &  \nabla_{\mu} \left( X^{q-1} \nabla^{\mu} \phi \right)  + \frac{\epsilon}{q f_1} \left( m f_0 \phi^{m-1} R^n - V_{\phi}  \right) = 0.
\end{eqnarray}
In this case the potential of the corresponding theory will be $V(\phi) = V_0 \phi^{ \frac{q (2n -m)}{n-q}}$ with $n \neq q$, and it is obtained the Noether symmetries for this potential such that ${\bf Y}_1, \, {\bf Y}_2$ given by (\ref{ns-12}) and
\begin{equation}
{\bf Y}_3 = \left[ \frac{m}{2 q} ( 3 - 2q ) + 2n -3 \right] A \partial_A + (1 - \frac{m}{2 q}) \left(  M \partial_M -  R \partial_R \right) - \frac{(n-q)}{2 q} \phi \partial_{\phi} + \frac{(m - 2n)}{2 q} X \partial_X,  \label{Y3-iv}
\end{equation}
which have the Noether integrals
\begin{eqnarray}
& & I_1 = - E_{\mathcal{L} }, \qquad I_2 = I_1 r + f_0 n R^{n-1} \phi^m M A F  \left[ (n-1) \frac{R'}{R} + \frac{M'}{M} \right],   \label{fint-iv-1} \\ & & I_3 =  M A F \left\{ f_0 n \phi^m R^{n-1} \left[ \ell \frac{A'}{A} + p \left( (n-1)\frac{R'}{R}  + m \frac{\phi'}{\phi} \right) \right] +  \frac{\epsilon}{2} f_1 (n - q) X^{q-1} \phi \phi'  \right\}, \label{fint-iv-2}
\end{eqnarray}
where $\ell$ and $p$ are defined as $\ell= 2 -n  + \frac{m}{2q} (q -2)$, $p= 2 n - 1 + \frac{m}{2 q}( 1 -2 q ), \, q \neq 0$, and $F = f_0 \phi^m R^{n-1} \left[ 2n + (n-1) M R \right] + f_1 (1-q) M X^q -  V_0 M \phi^{\frac{q (2n-m)}{n-q}}$.
The above relations give
\begin{eqnarray}
& & A (r) = \frac{ \left[ R^{n-1} \phi^m \right]^{-\frac{p}{\ell}}  }{ R_1(r) } \left[ A_0 + \frac{ I_3 }{f_0 \ell n } \int{  \frac{ \left[ R^{n-1} \phi^m \right]^{\frac{p-\ell}{\ell}} R_1 (r)} { M F } dr} \right],  \label{A-iv}  \\& &  B(r) = - \frac{I_1}{ A \, F^2},  \label{B-iv} \\& & (n-1) \frac{R'}{R} + \frac{M'}{M} = \frac{ (I_2 - I_1 r) R^{1-n} }{f_0 n \phi^m M A F},  \label{cnstra-iv}
\end{eqnarray}
where $A_0$ is an integration constant, $\ell \neq 0$,  and $R_1 (r) = \exp \left[  \frac{ \epsilon f_1 (n-q)}{2 f_0 \ell n}  \int{ \frac{ R^{1-n} \phi^{1-m} \phi' }{X^{1-q}} dr}  \right]$. It is seen that this case is a generalization of the previous case.

If $n= q =1$, i.e. $f(R,\phi, X) = f_0 R \phi^m + f_1 X - V(\phi)$, then we find the Noether symmetries ${\bf Y}_1, \, {\bf Y}_2$ given in \eqref{ns-12} and
\begin{equation}
{\bf Y}_3 = A \partial_A - M \partial_M ,  \label{Y3-iii-2}
\end{equation}
with vanishing potential $V(\phi) = 0$. Thus, the corresponding Noether integrals become
\begin{eqnarray}
& & I_1 = - E_{\mathcal{L} }, \qquad I_2 = I_1 r - I_1 f_0 M A \phi^m \left( \frac{M'}{M} + m \frac{\phi'}{\phi} \right),   \qquad  I_3 =  I_1 f_0 M A \phi^m \left( \frac{A'}{A} + m \frac{\phi'}{\phi} \right) , \label{fint-iii-2}
\end{eqnarray}
where $E_{\mathcal{L} } = 4 f_0^2 A B \phi^{ 2 m}$. These first integrals give rise to the solutions
\begin{eqnarray}
& & A = \phi^{-m} \left[ A_0 +  \frac{I_3}{I_1 f_0} \int{\frac{dr}{M} } \right],  \qquad B = - \frac{ I_1 \phi^{-2 m}}{4 f_0^2 A},  \label{sol-iii-AB} \\ & & \phi^m = \frac{1}{M} \left[  \phi_0 + \frac{1}{I_1 f_0} \int{ \frac{ (I_1 r + I_2)}{A} dr } \right],  \label{sol-iii-phi}
\end{eqnarray}
where $A_0$ and $\phi_0$ are integration constants. For the case $M=r^2$, assuming $\phi(r)=  \left[ \frac{ C_1 ( A_0 f_0 I_1 r-I_3 )}{r (I_1 r+I_2)}\right]^{1/m}$, it follows from \eqref{sol-iii-AB} and \eqref{sol-iii-phi} that there are the possibilities $(a)\,\, \phi_0  = 0,  A_0 = f_0^{-1}, I_2 = - I_3$ and $(b)\,\, I_2 = 0, A_0 = f_0^{-1}, I_1 = -  C_1 I_3 / \phi_0$. In case $(a)$, one gets the following analytical solutions
\begin{eqnarray}
A(r)= \frac{A_0}{C_1 I_1} \left( I_1 r - I_3 \right)\,, \quad B(r)= \frac{A_0 I_1^2 C_1^{1- 2 m} r^{ 2 m}}{4 (I_3 -I_1 r)}\,, \quad \phi(r)^m = \frac{C_1}{r}. \label{iii-sol1}
\end{eqnarray}
Then, in case $(b)$, the analytical solutions are found as follows
\begin{eqnarray}
A(r)= \frac{A_0}{C_1} r \,, \quad B(r)= -\frac{A_0 I_1 C_1 r }{4 \left( C_1 + \frac{\phi_0}{ r} \right)^2 }\,, \quad \phi(r)^m = \frac{C_1}{r} + \frac{\phi_0}{ r^2} \,, \label{iii-sol2}
\end{eqnarray}
where it should be $I_1 < 0$. To the best of our knowledge, these are also new spherically symmetric solutions in this non-minimally couple theory between the scalar field and the Ricci scalar. It should be noted that the solution \eqref{iii-sol2} only depends on $m$ in the scalar field and not in the metric. The metrics \eqref{iii-sol1} and \eqref{iii-sol2} are non-asymptotically flat, and they have horizons at $r=I_3/I_1$ and $r=0$, respectively.

\subsection{{\bf Case (iv):} $f(R,\phi,X)= f_0 R^n + f_1 \, \phi^m\, X^q$.\label{case4}}
Here, the field Eq.(\ref{fieldeq2}) and trace Eq.(\ref{trace}) become
\begin{eqnarray}
& & \square \left( R^{n-1} \right) + \frac{(n-2)}{3 n} R^n + \frac{(q-2) f_1}{3 n f_0} \phi^m X^q = 0,  \qquad  \nabla_{\mu} \left( \phi^m X^{q-1} \nabla^{\mu} \phi \right)  + \frac{\epsilon m}{q} \phi^{m-1} X^q  = 0.
\end{eqnarray}
For this case, we have the Noether symmetries ${\bf Y}_1, \, {\bf Y}_2$ and
\begin{eqnarray}
& & {\bf Y}_3 = (2n -3) A \partial_A + M \partial_M - R \partial_R - \frac{(n-q)}{(m+ 2 q)} \phi \partial_{\phi} - \frac{(m + 2n)}{(m + 2q)} X \partial_X,  \\& &  {\bf Y}_4 = \phi^{-\frac{m}{2q}} \left( \partial_{\phi} - \frac{m \, X}{ q\, \phi} \partial_X \right),   \label{Y3-v}
\end{eqnarray}
with $q \neq 0$. Then the first integrals corresponding to the above symmetries are
\begin{eqnarray}
& & I_1 = - E_{\mathcal{L} }, \quad I_2 = I_1 r + f_0 n M A F R^{n-1} \left[ (n-1) \frac{R'}{R} + \frac{M'}{M} \right],   \label{fint-v-1} \\ & & I_3 =  M A F \left\{ f_0 n R^{n-1} \left[ (2-n) \frac{A'}{A} + (n-1)(2 n -1) \frac{R'}{R} \right] + \frac{\epsilon f_1 (n-q)}{(m+ 2 q)}   X^{q-1} \phi^{m+1} \phi' \right\},  \label{fint-v-2}  \\& & I_4 = - \epsilon f_1 q M A F X^{q-1}  \phi^{ \frac{ m(2 q -1)}{2 q}}  \phi' \, , \quad \label{fint-v-3}
\end{eqnarray}
where $F = f_0 R^{n-1} \left[ 2 n + (1-n) M R \right] + f_1 (1-q) M \phi^m X^q$.
Then we solve the above equation (\ref{fint-v-2}) in terms of $A$ and find
\begin{eqnarray}
& &  A(r) = R^{ \frac{ (n-1)(2 n -1)}{n-2} } \left[ A_1  + \frac{1}{ f_0 n (2-n)} \int{  R^{ \frac{ (n-1)(4 n - 5)}{2-n}} \left( I_3 + \frac{ I_4 (n-q) \phi^{1 + \frac{m}{2q}} \phi'}{q (m + 2q) }  \right) \frac{dr }{ M F_1} } \right],
\end{eqnarray}
where $A_1$ is an integration constant, and $F_1 \equiv f_0 [ 2 n + (1-n) M R ] + f_1 (1-q) M R^{1-n} \phi^m X^q$. This solution reduces to the solution of case (i) for $A$ given in (\ref{i-A}) if $f_1 = 0$ and so $I_4 = 0$.

\subsection{{\bf Case (v):}  $f(R,\phi,X)=  U(\phi, X) \, R$.\label{case5}}
This case gives a Brans-Dicke type action, where the coupling to the Ricci curvature also includes the kinetic term of the scalar field $\phi$. Then we select some form of the function $U(\phi, X)$ to search the Noether symmetries.

\subsubsection{{\bf Subcase (v-a):} $U= f_0 X^q W(\phi)$.}
The field Eq.(\ref{fieldeq2}) and trace Eq.(\ref{trace}) in this subcase have the form
\begin{eqnarray}
& & R = \frac{ 3 X^{-q} }{ (q-1) W(\phi)} \square \left[ X^{q} W (\phi) \right],  \qquad  \nabla_{\mu} \left( X^{q-1} W R \nabla^{\mu} \phi \right)  + \frac{\epsilon }{q} X^{q} W_{\phi} R  = 0.
\end{eqnarray}
For this subcase, when $W(\phi)$ is an arbitrary function of $\phi$, it is found that the Noether symmetries are ${\bf Y}_1, {\bf Y}_2$ by (\ref{ns-12}) and
\begin{eqnarray}
& &  {\bf Y}_3 =  A \partial_A + \frac{1}{(2q -1)} \left( M \partial_M - R \partial_R -  X \partial_X \right), \quad q \neq \frac{1}{2} \, . \label{Y3-vi-a}
\end{eqnarray}
The corresponding first integrals of these symmetries become
\begin{eqnarray}
& & I_1 = - E_{\mathcal{L} }, \quad I_2 =  I_1 r + f_0^2 X^{2 q} W^2 M A  (2 - q M R) \left( \frac{M'}{M} + \frac{W'}{W} + q \frac{X'}{X} \right),  \label{fint-vi-a-1}  \\& & I_3 =  \frac{f_0^2}{ (2q- 1)}  X^{2 q} W^2 M A ( 2- q M R) \left[ (1-q) \frac{A'}{A} + (2q +1) \left( \frac{W'}{W} + q \frac{X'}{X} \right)  \right],  \label{fint-vi-a-2}
\end{eqnarray}
where $W' = W_{\phi} \phi'$. Solving the first integral (\ref{fint-vi-a-2}) in terms of $A$ it follows that
\begin{eqnarray}
& &  A =  \left( X^{ q} W \right)^{ \frac{ 2q +1}{q-1}} \left[  A_0 +  \frac{ (2 q -1) I_3}{ (1-q) f_0^2} \int{ \frac{ \left( X^{q} W \right)^{ \frac{ 4q -1 }{1-q}} dr}{ M ( 2 - q M R)} } \right] ,  \label{a-AB}
\end{eqnarray}
and combining (\ref{fint-vi-a-1}) yields
\begin{eqnarray}
& & B = - \frac{I_1}{ f_0^2 A X^{2 q} W^2 ( 2 - q MR)^2}, \label{fint-vi-a-1-2} \\& &
\frac{W'}{W} + \frac{M'}{M} +  q \frac{X'}{X}  = \frac{I_2 -  I_1 r}{f_0^2 X^{2 q} W^2 M A  (2 - q M R)} ,  \label{fint-vi-a-2-2}
\end{eqnarray}
where $A_0$ is a constant of integration. Considering these results one can derive some exact solutions of the field equations. As an example of the above solution, if we take $q= 1/4, M(r) = r^2$ and $R= \alpha/r^2$, then the relations (\ref{a-AB})-(\ref{fint-vi-a-2-2}) give
\begin{eqnarray}
& & A = \frac{A_0}{K_1} r^{3 - p} \left( K_1 \, r  + K_2 \right)^{1 + p \ell},  \label{vi-A} \\ & &
B = \frac{ 1 }{  K_1 + \frac{K_2}{r} }, \label{vi-BR}
\end{eqnarray}
where $p = \frac{ (8 -\alpha) I_2}{2 I_1 K_2}, \, \ell = \frac{ ( I_1 K_2 + I_2 K_1)}{ K_1 I_2}$, and $K_1, K_2$ are constants defined by $K_1 = - \frac{A_0 f_0^2 (8 - \alpha)^2}{16 I_1}$ and $K_2 = \frac{ (\alpha -8) I_3}{6 I_1}$. Considering the definition of $R$ given by (\ref{r-scalar}) to satisfy $R= \alpha/r^2$, we find that $ \ell = -1/3$ and $ K_1 = 1- \alpha/2$ for $p=3$ which means $A = \frac{A_0}{K_1}$ and $B$ by (\ref{vi-BR}), or $ \ell = -1/4$ and $ K_1 = (4- 2 \alpha)/3$ for $p=4$ which means $A = \frac{A_0}{K_1 r}$ and $B$ by (\ref{vi-BR}).

\bigskip

\subsubsection{{\bf Subcase (v-b):} $U= f_0 X^{1/2} V_{\phi}$, where $V_{\phi} = d V(\phi) / d\phi$.}
For this subcase, the field Eq.(\ref{fieldeq2}) and trace Eq.(\ref{trace}) are as follows
\begin{eqnarray}
& & R = - \frac{ 6 }{ V_{\phi} \sqrt{X}} \square \left( V_{\phi} \sqrt{X} \right),  \qquad  \nabla_{\mu} \left( \frac{R}{ \sqrt{X}} V_{\phi} \nabla^{\mu} \phi \right)  + 2 \epsilon \sqrt{X} V_{\phi \phi} R  = 0.
\end{eqnarray}
This subcase has also Noether symmetries  ${\bf Y}_1, {\bf Y}_2$ given in (\ref{ns-12}) and  ${\bf Y}_3$ with
\begin{eqnarray}
& &  {\bf Y}_3 =  M \partial_M - R \partial_R -  X \partial_X . \label{Y3-vi-b}
\end{eqnarray}
In this subcase, we find some extra Noether symmetries as follows
\begin{eqnarray}
& & {\bf Y}_4 = A \partial_A - \frac{ (M R -4)}{2 M} \partial_R - \frac{V}{M R V_{\phi}} \partial_{\phi} +  \left( \frac{2 V V_{\phi \phi}}{M R V_{\phi}^2} - \frac{1}{2} \right) X \partial_X, \label{iv-Y4} \\& &  {\bf Y}_5 =  - \frac{ (M R -4)}{M} \partial_R + \frac{ (M R -2) V}{M R V_{\phi}} \partial_{\phi} + \left[ 1 + \frac{2 ( 2 -M R) V V_{\phi \phi}}{M R V_{\phi}^2} \right] X \partial_X, \label{iv-Y5} \\& & {\bf Y}_6 = A \ln M \partial_A - \frac{M}{2} \ln \left( A M^3 \right) \partial_M  + \frac{2}{M} \left[ (M R +1) \ln M + \ln A \right] \partial_R   - \frac{V}{M R V_{\phi}} \left[ (M R -1) \ln M + \left( 1 - \frac{M R}{2} \right) \ln A  \right] \partial_{\phi} \nonumber \\& & \quad \qquad + \left[ 1 + \ln A + \frac{2 V V_{\phi \phi}}{M R V_{\phi}^2} \left( (M R -1) \ln M + \left( 1 - \frac{M R}{2} \right) \ln A  \right) \right] X \partial_X, \label{iv-Y6} \\& & {\bf Y}_7 = A \ln A \left( \ln M  + \frac{\ln A}{2} \right) \partial_A + M \ln M \left( \ln M +  \frac{\ln A}{2} \right) \partial_M  \nonumber \\& & \qquad + \frac{1}{M} \left\{ \left[ \left(\frac{3 M R}{2} - 2 \right) \ln M + (M R -2) \ln A 2 M R -8 \right] \ln M + \left[ \left( \frac{M R}{2} -2 \right)  \ln A +  M R -4 \right] \ln A \right\} \partial_R  \nonumber \\& & \qquad  + \frac{V}{M R V_{\phi}} \left\{ \left[ \left( \frac{M R}{2} - 1 \right) \ln M + 2 M R -4 - \ln A \right] \ln M + \left[ \left( \frac{M R}{4} -1 \right)  \ln A +  M R -2 \right] \ln A \right\} \partial_{\phi} \nonumber \\& & \qquad + \Big[ \frac{2 V V_{\phi \phi}}{M R V_{\phi}^2} \left\{ \left[ \left( 1- \frac{M R}{2} \right) \ln M + \ln A - 2 M R + 4 \right] \ln M +  \left[  \left( 1 - \frac{M R}{4} \right) \ln A  - M R + 2 \right] \ln A  \right\}  \nonumber \\& & \qquad \qquad - \left( \ln A + \frac{\ln M}{2} \right) \ln M \Big] X \partial_X, \label{iv-Y7}
\end{eqnarray}
\begin{eqnarray}
& & {\bf Y}_8 = A \ln A \partial_A  + 2 M \left( \ln M + \frac{M R}{4} \right) \partial_M - \frac{1}{M} \left[ ( 3 M R - 4) \ln M + (M R -2 )\ln A \right] \partial_R  \nonumber \\& & \qquad + \frac{V}{M R V_{\phi}} \left[ (M R -2) \ln M - \ln A  \right] \partial_{\phi} + \left[ \frac{2 V V_{\phi \phi}}{M R V_{\phi}^2} \left( ( 2 - M R) \ln M + \ln A \right) - \ln (M A) -2 \right] X \partial_X, \\& & {\bf Y}_9 = V \left[ A \partial_A + M \partial_M - R \partial_R + \left( \frac{3 \epsilon X V_{\phi}}{ R V} - \frac{V}{4 V_{\phi}} \right) \partial_{\phi} -  \left( 2 +  \frac{6 \epsilon X V_{\phi \phi}}{ R V} - \frac{V V_{\phi \phi}}{ 2 V_{\phi}^2}  \right) X \partial_X \right] , \\& & {\bf Y}_{10} = \frac{4 \epsilon r}{ f_0^2 M A R ( M R -4) V_{\phi}} \left[ \partial_{\phi} - \frac{2 V_{\phi \phi}}{ V_{\phi}}  X \partial_X \right] \qquad {\rm with} \quad K = V(\phi),
\end{eqnarray}
where the subscript $\phi$ denotes the derivative with respect to $\phi$, and $V_{\phi} \neq 0$. The Noether symmetries ${\bf Y}_1, ..., {\bf Y}_9$ have the following nonvanishing Lie brackets:
\begin{eqnarray}
& & \left[ {\bf Y}_1, {\bf Y}_2 \right] =  {\bf Y}_1, \qquad \left[ {\bf Y}_2, {\bf Y}_6 \right] = -\frac{1}{2} {\bf Y}_3 + \frac{1}{2} {\bf Y}_5, \qquad \left[ {\bf Y}_2, {\bf Y}_7 \right] = {\bf Y}_5 + {\bf Y}_6 + {\bf Y}_8, \qquad \left[ {\bf Y}_2, {\bf Y}_8 \right] = \frac{1}{2} {\bf Y}_3 + {\bf Y}_4, \nonumber \\& &  \left[ {\bf Y}_3, {\bf Y}_6 \right] =- \frac{3}{2} {\bf Y}_3 + {\bf Y}_4 - {\bf Y}_5 , \qquad \, \left[ {\bf Y}_3, {\bf Y}_7 \right] = 2 {\bf Y}_5 + {\bf Y}_8, \qquad \qquad \left[ {\bf Y}_3, {\bf Y}_8 \right] = 2 {\bf Y}_3 + {\bf Y}_5 , \label{nsv-v-b-Lb}    \\& & \left[ {\bf Y}_4, {\bf Y}_6 \right] = -\frac{1}{2} {\bf Y}_3 + \frac{1}{2} {\bf Y}_5 ,  \qquad \qquad  \left[ {\bf Y}_4, {\bf Y}_7 \right] = {\bf Y}_5 + {\bf Y}_6 + {\bf Y}_8, \qquad \left[ {\bf Y}_4, {\bf Y}_8 \right] = \frac{1}{2} {\bf Y}_3 + {\bf Y}_4 , \nonumber \\& &  \left[ {\bf Y}_6, {\bf Y}_7 \right] = -{\bf Y}_7, \quad \left[ {\bf Y}_6, {\bf Y}_8 \right] = - {\bf Y}_6 - \frac{1}{2} {\bf Y}_8  , \qquad \left[ {\bf Y}_7, {\bf Y}_8 \right] = -2 {\bf Y}_7 . \nonumber
\end{eqnarray}
Here we do not consider the Lie brackets of Noether symmetry ${\bf Y}_{10}$ due to the gauge function $K = V(\phi)$ appeared together with this symmetry. For the Noether symmetries ${\bf Y}_{1}$ and ${\bf Y}_{10}$, the Noether first integrals are
\begin{equation}
I_1 = - E_{\mathcal{L} }, \qquad I_{10} = r V_{\phi} \phi' - V(\phi),  \label{fint-vi-b-1}
\end{equation}
in which the latter first integral has a solution for $V(\phi)$ in terms of $r$ as $V(\phi) = V_0 r - V_1$, where $V_1 \equiv I_{10}$ and $V_0$ is an integration constant. Thus, using the definition of $X$ which has the form $X = - \epsilon \phi'^2 / 2 B$, we find from the first relation of (\ref{fint-vi-b-1}) for $A$ that
\begin{eqnarray}
& & A = \frac{ 8 \epsilon I_1}{ f_0^2 V_{0}^2 ( 4- M R)^2  }.  \label{v-b-A}
\end{eqnarray}
For the first integrals of the Noether symmetries ${\bf Y}_2,..., {\bf Y}_9$, one can get the following relations
\begin{eqnarray}
& & I_2 =  I_1 r + \frac{2 I_1 M }{ B (4 - M R) } \left( \frac{M'}{M} - \frac{B'}{2 B} \right),   \quad I_3 =  \frac{2 I_1 M }{ B (4 - M R)} \left( \frac{A'}{2A} -  \frac{B'}{B} \right), \nonumber \\& & I_4 = \frac{ I_1 M}{ B ( 4- M R)} \left( \frac{A'}{2 A} + \frac{B'}{B} - \frac{M'}{M} +  \frac{2 V B}{V_0 M} \right), \qquad  I_5 = \frac{- 2 I_1 M}{ B ( 4- M R)} \left[ \frac{A'}{ A} + \frac{2 M'}{M} +  \frac{2 V B }{V_0 M} (MR -2) \right], \nonumber \\ & & I_6 = \frac{ - I_1 M}{ B ( 4- M R)} \left[ \ln A \left( \frac{B'}{B} + \frac{M'}{M} +  \frac{V B}{V_0 M} (M R -2) \right) + \ln M \left( \frac{2 B'}{ B} - \frac{3 A'}{A} - \frac{M'}{M} - \frac{ 2 V B}{V_0 M} (M R -1) \right) +  \frac{ A'}{A} +  \frac{2 M'}{M} \right], \nonumber \\ & & I_7 = \frac{ - I_1 M}{ B ( 4- M R)} \Big[ (\ln A)^2  \left( \frac{M'}{M} -\frac{B'}{2B}   + \frac{V B}{2 V_0 } (M R -4) \right) + 2 \ln A \ln M \left( \frac{M'}{2 M} - \frac{B'}{B} - \frac{V B}{V_0} \right)  \nonumber \\& & \qquad \qquad \qquad \qquad \quad  + 2 (\ln M)^2 \left( \frac{3 A'}{2 A} - \frac{B'}{B} + \frac{M'}{2 M}  + \frac{ V B }{2 V_0} (M R -2) \right) + 2 \ln (A M^2) \frac{ V B}{V_0} (M R -2) \Big], \label{fint-1-9} \\ & & I_8 = \frac{ 2 I_1 M}{ B ( 4- M R)} \Big[  \ln A \left( \frac{A'}{A} + \frac{B'}{B} + \frac{V B}{V_0 M} \right) - \ln M \left(  \frac{3 A'}{A} - \frac{2 B'}{B} + \frac{M'}{M} + \frac{V B}{V_0 M} (M R -2) \right) \nonumber \\& & \qquad \qquad \qquad \qquad \quad   - \frac{M R}{2} \left( \frac{A'}{A} - \frac{B'}{B} + \frac{M'}{M} \right) + \frac{A'}{A} - \frac{2 M'}{M} \Big], \nonumber \\ & & I_9 = \frac{ I_1 M}{ B ( 4- M R)} \left[ \frac{3 B'}{B} + \frac{3 V_0}{V} + \frac{ V B R}{2 V_0} \right], \nonumber
\end{eqnarray}
with $V= V_0 r - V_1$. Then, it is obtained for $B$ from the first integrals $I_2, I_3, I_4$ and $I_9$ that
\begin{eqnarray}
& &  B = \frac{ 3 V_0 I_1 M  }{ \left[ K_0 +  \frac{I_1 (V_0 r - V_1)}{2 V_0} \right] M R - 4 K_0 }, \label{v-b-B}
\end{eqnarray}
where $K_0$ is a constant defined by $K_0 = V_1 \frac{I_1}{V_0} - I_2 + I_3 + 2 I_4 - I_9$. Thus, the first integrals $I_2$ and $I_3$ in (\ref{fint-1-9}) yield
\begin{eqnarray}
\left[ K_0 +  \frac{I_1 (V_0 r - V_1)}{2 V_0} \right] M R - 4 K_0 & = &\frac{ 3 V_0}{M} \left(  B_0 I_1 - \int{ (I_1 r - I_2) (4 - M R) M dr } \right), \label{ceq-v-b-1} \\& = & 3 V_0 M ( 4 - M R ) \left( B_1 I_1 - \frac{I_3}{2} \int{ \frac{dr}{M}} \right), \label{ceq-v-b-2}
\end{eqnarray}
where $B_0, B_1$ are constants of integration.
Furthermore, after some algebra, we find from the first integrals (\ref{fint-1-9}) the following constraint relations
\begin{eqnarray}
& & I_5 = 4 \left( I_2 - I_4 - \frac{V_1}{V_0} I_1 \right), \nonumber \\& & I_6 = \frac{I_5}{2} +  \frac{ 2 I_1 V (M R -2) }{ V_0 ( 4 - M R)} +  \left[ \frac{ 2 I_1 M }{ B ( 4 - M R)} \left( \frac{V_0}{V} + \frac{V B R}{6 V_0} \right) - I_3 - \frac{I_5}{2} - \frac{4}{3} I_9 \right] \ln A \nonumber \\& & \qquad \quad + \left[ \frac{ 2 I_1 V (M R -1) }{ V_0 ( 4 - M R)}- \frac{ 5 I_1 M }{ B ( 4 - M R)}  \left( \frac{V_0}{V} + \frac{V B R}{6 V_0} \right) + I_1 r - I_2 + 3 I_3 + \frac{5}{3} I_9 \right] \ln M, \nonumber \\& & I_7 = -\frac{ 2 I_1 V (M R -2) }{ V_0 ( 4 - M R)} \ln (A M^2) + \left[ \frac{ I_1 V M (M R -2) }{ V_0 ( 4 - M R)}  - \frac{ I_1 M }{ 2 B ( 4 - M R)} \left( \frac{V_0}{V} + \frac{V B R}{6 V_0} \right) - I_1 r + I_2 - \frac{I_9}{6} \right] (\ln A)^2 \nonumber \\& &  \qquad \quad - \left[ \frac{ I_1 M }{ B ( 4 - M R)} \left( \frac{V_0}{V} + \frac{V B R}{6 V_0} \right) - \frac{ 2 I_1 V }{ V_0 ( 4 - M R)} + I_1 r - I_2 - \frac{I_9}{3} \right] (\ln A) ( \ln M) \nonumber \\& & \qquad \quad + \left[ \frac{ 5 I_1 M }{ B ( 4 - M R)} \left( \frac{V_0}{V} + \frac{V B R}{6 V_0} \right) - \frac{ I_1 V M (M R -2) }{ V_0 ( 4 - M R)} - I_1 r + I_2 - 3 I_3- \frac{5}{3} I_9  \right] (\ln M)^2, \label{fint-1-9-2} \\& & I_8 = 2 I_3 - (I_1 r -I_2) (M R + 4) + \left[  \frac{ 2 I_1 M }{ B ( 4 - M R)} \left( \frac{V_0}{V} + \frac{V B R}{6 V_0} \right) - I_3 - \frac{2}{3} I_9 \right] M R \nonumber \\& & \qquad \quad + 2 \left[ I_3 + I_9 + \frac{ I_1 V }{ V_0 ( 4 - M R)} - \frac{ 3 I_1 M }{ B ( 4 - M R)} \left( \frac{V_0}{V} + \frac{V B R}{6 V_0} \right) \right] \ln A \nonumber \\& & \qquad \quad - 2 \left[ \frac{ I_1 V (M R -2) }{ V_0 ( 4 - M R)} - \frac{ 5 I_1 M }{ B ( 4 - M R)}  \left( \frac{V_0}{V} + \frac{V B R}{6 V_0} \right) + I_1 r - I_2 + 3 I_3 + \frac{5}{3} I_9 \right] \ln M. \nonumber
\end{eqnarray}
Now, for $M = r^2$, the constraint equation (\ref{ceq-v-b-2}) yields
\begin{equation}
R = \frac{ 8 \left[ 3 V_0 \left( B_1 I_1 + \frac{I_3}{2 r} \right) + \frac{K_0}{r^2}  \right] }{ 2 K_0 - \frac{V_1 I_1}{V_0} + ( I_1 + 3 V_0 I_3) r + 6 V_0 B_1 I_1 r^2}, \label{v-b-R2}
\end{equation}
Then, using the latter $R$ in Eqs. (\ref{v-b-A}) and (\ref{v-b-B}) for $A$ and $B$, respectively, it follows that
\begin{eqnarray}
& & A = \frac{ \epsilon \left[ 2 K_0 - \frac{V_1 I_1}{V_0} + ( I_1 + 3 V_0 I_3) r + 6 V_0 B_1 I_1 r^2 \right]^2}{ 2 I_1 f_0^2 ( V_0 r - V_1)^2 }, \label{v-b-A2}  \\& & B = \frac{ V_0 \left[ 2 K_0 - \frac{V_1 I_1}{V_0} + ( I_1 + 3 V_0 I_3) r + 6 V_0 B_1 I_1 r^2  \right]}{ 4 (V_0 r - V_1)\left( B_1 I_1 + \frac{I_3}{2 r} \right)}, \label{v-b-B2}
\end{eqnarray}
which is a new solution for $M = r^2$. Some other solutions could be produced from the constraint equation (\ref{ceq-v-b-2}) if we choose an integrable function of $M$ in it. Clearly, this solution is non-asymptotically flat unless $I_1=-3 V_0 I_3$ and $B_1=0$. The singularities of the solutions \eqref{v-b-A2} and \eqref{v-b-B2} can be classified in the following way:
\begin{itemize}
\item At $r = V_1 / V_0$, the metric coefficients $A$ and $B$ tend to infinity, but the Ricci scalar remains finite there.

\item At $r=-I_3/(2B_1 I_1)$, the metric coefficient $A$ becomes finite, $B$ goes to infinity, and the Ricci scalar remains finite there.
    
\item The algebraic equation $A(r) = 0$ from \eqref{v-b-A2} gives the solutions
\begin{eqnarray}
& & r_{\pm} = \frac{1}{ 12 V_0 B_1 I_1} \left[ -( I_1 + 3 V_0 I_3) \pm \sqrt{ ( 24 B_1 V_1 + 1) I_1^2 + 6 V_0 I_1 ( I_3 - 8 B_1 K_0 ) + 9 V_0^2 I_3^2 }  \right].
\end{eqnarray}
Here, the outer horizon $r= r_{+}$ of the metric constructed from \eqref{v-b-A2}, \eqref{v-b-B2} and $M(r)= r^2$ exhibits an event horizon. The inner horizon $r = r_{-}$  is not
because in the entire region $r < r_{+}$ there are outgoing radial null geodesics which
fail to reach future null infinity and the hypersurface $r = r_{-}$ is not a boundary of a region
with this property. At $r=r_{\pm}$, the metric functions $A$ and $B$ vanish, but the Ricci scalar diverge to infinity, which is a true space-time singularity. 
\end{itemize}

\section{Conclusions}

In this paper we derived the Noether symmetries of spherically symmetric metric (\ref{metric}) for a Lagrangian density with the function $f(R, \phi, X)$. This analysis covers most of modified gravity models proposed in the current literature.  It is important to get any exact solutions for a given theory admitting a Noether symmetry if it exists. Besides, the existence of a Noether symmetry “select" the integrable form of a model in a given class of theories. Furthermore, the existence of Noether symmetries means to find out conserved quantities according to the Noether Theorem. For each form of the function $f(R,\phi,X)$, and so for the theory of gravity, we can find out exact cosmological solutions if there exists any Noether symmetry.

One can search for symmetries of the Lagrangian related to cyclic variables to reduce the dynamics. It is known that the conserved quantities are related to the existence of cyclic variables into the dynamics by the Noether symmetry (see Ref.~\cite{capo2007} for details). But it is not unique to find those of cyclic variable because of that the required equations for the change of coordinates have not unique solution, and it is usually needed a clever choice. Also, the solution of equations for the choice of coordinates is not well defined on the whole space \cite{capo2007}. Throughout this study, we deduced that it is better to use the classical Noether symmetry approach to find Noether symmetry in $f(R,\phi,X)$ theory of gravity, rather that the approach used in Ref. \cite{capo2007}. In this study we show that under the classical Noether theorem, Noether symmetry in $f(R,\phi,X)$ theory of gravity yields a rather handy conserved quantity (or the first integral of motion), which can be solved easily, and it is not required to search for the cyclic coordinate. Therefore, we directly use the conservation relation (\ref{con-law}) associated with the obtained Noether symmetry ${\bf Y}$ in order to find exact solutions for the field equations associated with the Lagrangian (\ref{lagr2}).
In the previous section, we have studied different kinds of $f(R,\phi,X)$ theories, in all of which the Noether symmetry exists and find exact spherically symmetric solutions in the corresponding $f(R,\phi,X)$ theory gravity. It is important to remark that in the literature, many authors have found solutions in $f(R)$ only considering $R=\textrm{constant}$, which indeed is a trivial case since all the higher order terms considered in $f(R)$ disappears. We have found new spherically symmetric solutions in power-law $f(R)$ gravity by considering $R\neq \textrm{constant}$ which is non-trivial and in principle, without using Noether symmetries, it could have been hard to find them. In addition, we have also found other solutions in theories considering the scalar field $\phi$, a kinetic term $X$ and a potential $V(\phi)$. Some interesting new spherically symmetric solutions were found for non-minimally couplings theories between the scalar curvature and the scalar field $f(R,\phi,X)=f_0 R^n \phi^m+f_1 X^q-V(\phi)$, non-minimally couplings between the scalar field and a kinetic term $f(R,\phi,X)=f_0 R^n +f_1\phi^mX^q$ , and also in extended Brans-Dicke gravity $f(R,\phi,X)=U(\phi,X)R$. Some of these solutions also represent black hole solutions. For some class of gravity theories in this paper we found some selected potential functions of $V(\phi)$. So one can apply the Noether symmetry approach as a selection rule to determine the form of the potential function $V(\phi)$ of the theory.

\begin{acknowledgments}

S.B. is supported by the Comisi{\'o}n Nacional de Investigaci{\'o}n Cient{\'{\i}}fica y Tecnol{\'o}gica (Becas Chile Grant No.~72150066) and Mobilitas Pluss N$^\circ$ MOBJD423 by the Estonian government. In addition, the work of K.B. is supported in part by the JSPS KAKENHI Grant Number JP25800136 and Competitive Research Funds for Fukushima University Faculty (18RI009). The authors are thankful to the referee for his/her valuable comments to improve our manuscript.

\end{acknowledgments}

%\bibliographystyle{Style}
%\bibliography{bibtele}

\end{document}